\documentclass[12pt]{article}
\usepackage{epsfig}
  \newcommand{\name}[1]{}               
\usepackage[usenames]{color}
\newcommand{\added}[1]{{\color{black}#1}}
\newcommand{\removed}[1]{}
\newcommand{\change}[2]{\added{#2}}
\newcommand{\addedd}[1]{{\color{black}#1}}
\newcommand{\removedd}[1]{}

\begin{document}
\title{MHD and Gyro-kinetic Stability of JET Pedestals}

\pagebreak
\maketitle
\begin{center}
S. Saarelma\dag, M.N.A. Beurskens\dag, D. Dickinson\dag,
L. Frassinetti\S, M.J. Leyland\ddag, C.M. Roach\dag, EFDA-JET Contributors
\footnote{See the Appendix of F. Romanelli et
  al., Proceedings of the 24th IAEA Fusion Energy Conference 2012, San Diego, USA}
\\
JET-EFDA, Culham Science Centre, Abingdon, OX14 3DB, UK\\
\dag Euratom/CCFE Fusion Association, Culham Science Centre, OX14
  3DB, Abingdon, UK\\
\ddag York Plasma Institute, Dept. of Physics, University of York,
  York, YO10 5DD, UK\\
\S Division of Fusion Plasma Physics, School of Electrical
Engineering, Royal Institute of Technology, Association EURATOM-VR, Stockholm,
Sweden\\
\end{center}
samuli.saarelma@ccfe.ac.uk

\section*{Abstract}
The pedestal profile measurements in high triangularity JET plasmas show
that with low fuelling the pedestal width decreases during the ELM cycle
and with high fuelling it stays constant. In the low fuelling case the
pedestal pressure gradient keeps increasing until the ELM crash and in the
\change{low}{high} fuelling case it \change{reaches a saturation}{initially increases then saturates} during the ELM cycle.

\added{Stability analysis reveals that both JET plasmas become unstable to finite-$n$ ideal 
MHD peeling-ballooning modes at the end of the ELM cycle. During the ELM cycle, $n=\infty$ ideal MHD 
ballooning modes and kinetic ballooning modes are found to be locally stable in
most of the steep pressure gradient region of the pedestal owing to the large bootstrap current, 
but to be locally unstable in a narrow region of plasma at the extreme edge.}

Unstable micro-tearing modes are found at the JET pedestal top, but they
are sub-dominant to ion temperature gradient modes. They are insensitive to
collisionality and stabilised by increasing density gradient.

\section{Introduction}
The global plasma confinement in a tokamak operating in high confinement or
H-mode is largely determined by the edge pedestal pressure due to core
turbulence restricting $\nabla T/T$ near the marginal stability
limit\cite{ryter}. Therefore, being able to predict the edge pedestal
behaviour allows prediction of the achievable core pressure and the fusion
power. There is strong evidence from various devices that in Type I ELMy
H-mode the pedestal pressure is constrained by ideal MHD peeling-ballooning
modes. When the stability limit is exceeded, the peeling-ballooning mode is
thought to trigger an instability causing an ELM crash that reduces the
pedestal top pressure. The stability calculations of JET \cite{saarelma1},
DIII-D \cite{snyder}, JT-60U \cite{jt60stability} and ASDEX Upgrade
\cite{snyder} have all found plasma to be within error margin of the
peeling-ballooning stability limit prior to a Type I ELM crash.

However, the peeling-ballooning constraint does not solely determine the
pedestal pressure. By increasing the width of the pedestal it is possible
to maintain peeling-ballooning stability at higher pedestal top pressure
with a reduced pressure gradient. In order to uniquely predict the maximum
achievable pedestal \added{height}, the pedestal width also has to be
constrained. \added{The EPED1 model uses a width scaling derived
from the physics of kinetic ballooning modes, $\Delta_{ped}=0.076\beta_{p,ped}^{0.5}$, 
combined with the peeling-ballooning stability limit to determine
the height of the pedestal \cite{snyderpop09}. Pedestal measurements from ASDEX-Upgrade, 
DIII-D, and JET have been shown to be consistent with EPED1 model predictions
\cite{beurskens}.} \addedd{A more
advanced version of EPED, EPED 1.6, replaces the scaling law constraint
with a procedure, outlined in \cite{eped16}, that extrapolates from local ideal MHD
stability calculations for $n = \infty$ ballooning modes to approximate the
stability of kinetic ballooning modes (KBM).} Both the
linear growth rate and the non-linear heat flux from KBMs increase very
rapidly after the stability limit is exceeded, constraining the pedestal
pressure gradient below this limit \cite{snyder01}. \added{ While the
local calculation in the pedestal region may not be sufficient to solve the
turbulent heat fluxes due to radial requirements of the computational box,
linear analysis can provide the local stability boundary.}

In recent gyrokinetic studies of MAST plasmas, in addition to KBMs,
micro-tearing modes (MTM) were found unstable near the top of the pedestal
\cite{dickinsonppcf,dickinsonprl}. This suggests a model where the widening
of the pedestal is due to MTMs in the shallow gradient region at the top of
the pedestal becoming stabilised by increasing density and pressure
gradients at the ``knee'' of the pedestal top, allowing the steep pressure
gradient to increase until the KBM stability limit is reached
\cite{dickinsonprl}.

Whilst the EPED model has been able to predict a large number of JET
pedestal pressures at a 20-30\% accuracy \cite{beurskens2}, some of the
trends are not correctly predicted. In this paper we investigate the linear
stability limit in the pedestal region of well-diagnosed ELM cycles in
JET. \addedd{The selected plasmas are
from a density scan that shows mixed agreement with EPED predictions
\cite{leyland}. The pedestal width increases with $\beta_{p,ped}$ as predicted by the
EPED model, but the scaling exponent cannot be determined accurately from
the data. The EPED predicted pedestal pressure, however, shows a
decreasing trend with increasing density while the opposite is observed
experimentally: this discrepancy is discussed in more detail in \cite{leyland}.}

We try to understand the underlying causes for the discrepancies by using
linear ideal MHD and \added{local} gyrokinetic stability analyses. First in section 2, we
construct the plasma equilibrium using the HELENA code \cite{HELENA}
including the bootstrap current that is dominant in the edge region. HELENA
also calculates the local $n=\infty$ ballooning stability for each flux
surface. Then we use the HELENA equilibria to study the peeling-ballooning
stability with a finite-$n$ MHD stability code MISHKA-1 in section
3. Finally, the gyrokinetic code GS2 \cite{gs2} is used to investigate the
micro-stability in section 4. We determine the 
\change{KBM unstable region}{region that is locally unstable to KBMs}
during
the ELM cycle and how well this corresponds to the region that is unstable to
the ideal MHD $n=\infty$ ballooning modes. We will also investigate, whether
unstable MTMs are found near the \added{JET} pedestal top.

\section{Experiment and Equilibrium Reconstruction}
\subsection{Equilibrium Reconstruction Procedure}
In order to study the MHD and gyrokinetic stability of the JET pedestal, we
first need to accurately reconstruct the equilibrium. The electron density
and temperature are obtained using the high resolution Thomson scattering
diagnostics \cite{pasquolotto,frassinetti}. The profiles are measured in
$R,Z$ coordinates, mapped to equilibrium poloidal flux by using the EFIT
equilibrium, binned according to their timing in the ELM cycle and then the
combined profiles are fitted with a $mtanh$-function \cite{mtanh} using a
deconvolution technique described in \cite{frassinetti}. Since the EFIT
separatrix position is not very accurate, we adjust the radial position of
the profiles so that the separatrix temperature matches with two-point
power balance model\cite{akallen}:
\begin{equation}
T_{e,sep} [eV]=\left(T_{div}^{\frac{7}{2}}+\frac{7}{2}(P_{heat}
 -P_{rad})\frac{L_{||}}{\lambda_q
  2\pi R_{OMP} k_0}\right)^{\frac{2}{7}}
\end{equation}
where $T_{div}$ is the temperature at the divertor (in eV), $P_{heat}$ is the total
heating power (in MW), $P_{rad}$ is the total radiative power from the core
(in MW), $L_{||}=\pi R_0 q_{95}$ is the parallel connection length (in m), $\lambda_q$ is
the radial power decay length (in m), $R_{OMP}$ is the major radius of the outer
midplane (in m) and $k_0$ is the heat conduction. The values used in this paper
are either taken from the experiment ($R_0, P_{heat}, P_{rad}, q_{95},
R_{OMP}$) or assumed ($\lambda_q=0.005$, $k_0=2000$). 

For ion temperature, we have assumed that in the core $T_i=T_e$ and used
two different assumptions for the pedestal, either $T_i=T_e$ or
$dT_{i,ped}/d\psi=dT_{i,core}/d\psi$. It turns out that the choice of the
assumption of ion temperature has little effect on the equilibrium and
stability. For the $Z_{\rm eff}$, which affects the bootstrap current through
collisionality as well as ion density through dilution, we have used a
line-integrated measurement and assumed constant value in the
plasma. \added{ The line integrated $Z_{\rm eff}$ is taken from the
visible Bremsstrahlung diagnostic.}

The current profile is assumed to be a combination of inductively driven
and bootstrap current. \added{ In the absence of edge current
measurements} we calculate the bootstrap current from density and
temperature profiles using formulas in
\cite{sauter,sautercorrection}. \added{ We note that the bootstrap calculation 
in this paper assumes constant $Z_{\rm eff}$ across the plasma, and that more recent
improved calculations predict lower bootstrap current density at high collisionality
\cite{Koh}. In the
stability analysis part we will conduct a scan in bootstrap current to
evaluate the sensitivity of the stability results due to the variation in
the bootstrap current.} The inductively driven current is assumed to be
fully relaxed to the neo-classical conductivity profile, which slightly
overestimates the core current but has little effect on the equilibrium
near the edge where the bootstrap current dominates. The plasma boundary
shape is taken from EFIT. Using this information, we calculate a fixed
boundary equilibrium that has a self-consistent bootstrap current with the
HELENA code using the method described in \cite{ssJETsmallELMs}.

To improve further the quality of the equilibrium, we redo the mapping of
the profiles from real space into the flux space using the HELENA
equilibrium. It turns out that the effect of this remapping on the profiles
is small and the edge pressure gradient changes by less than 5\%.

\subsection{Investigated Plasmas}
\begin{figure}[!b]
\begin{minipage}{.5\linewidth}
\epsfig{file=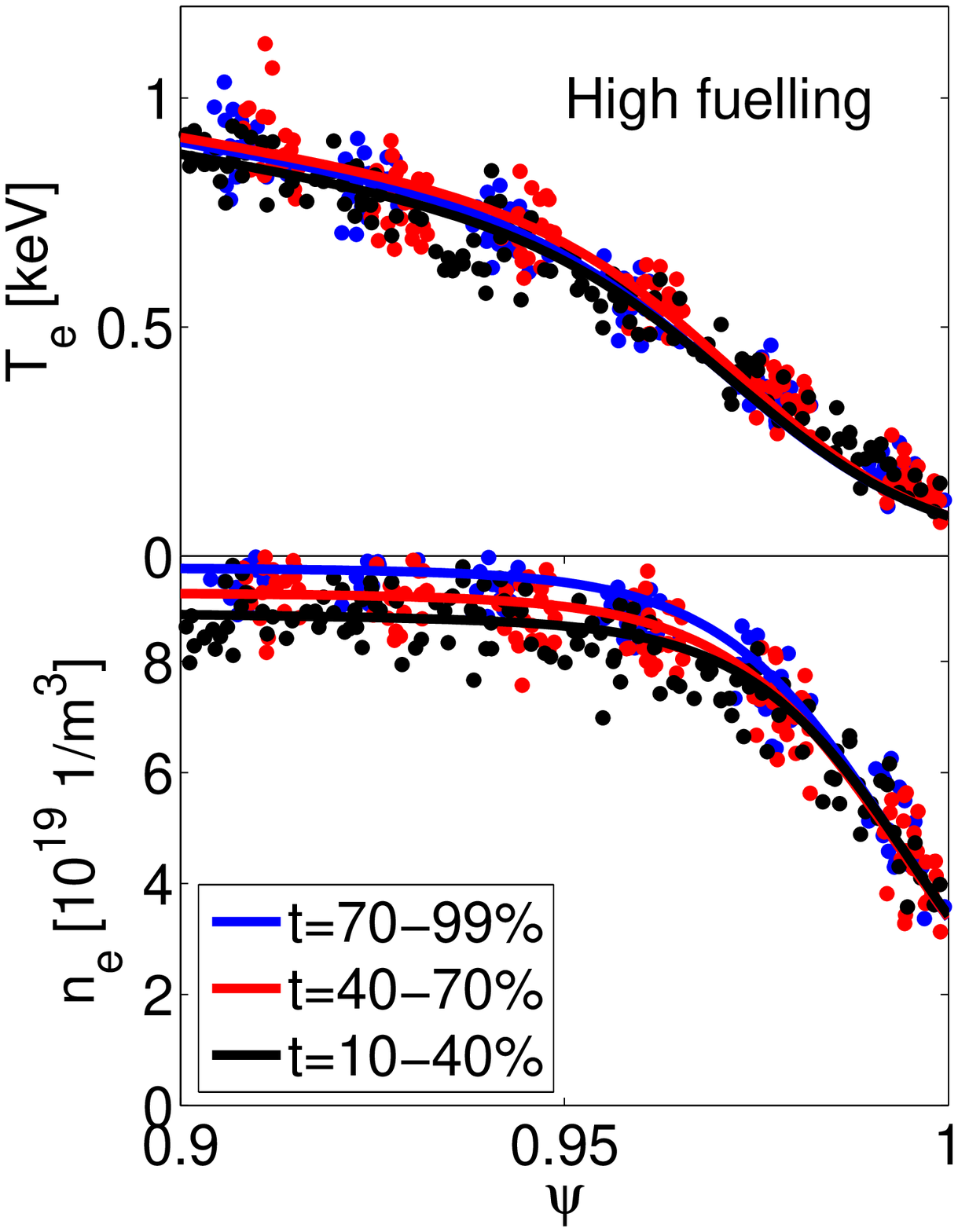,width=\linewidth}
\end{minipage}
\begin{minipage}{.5\linewidth}
\epsfig{file=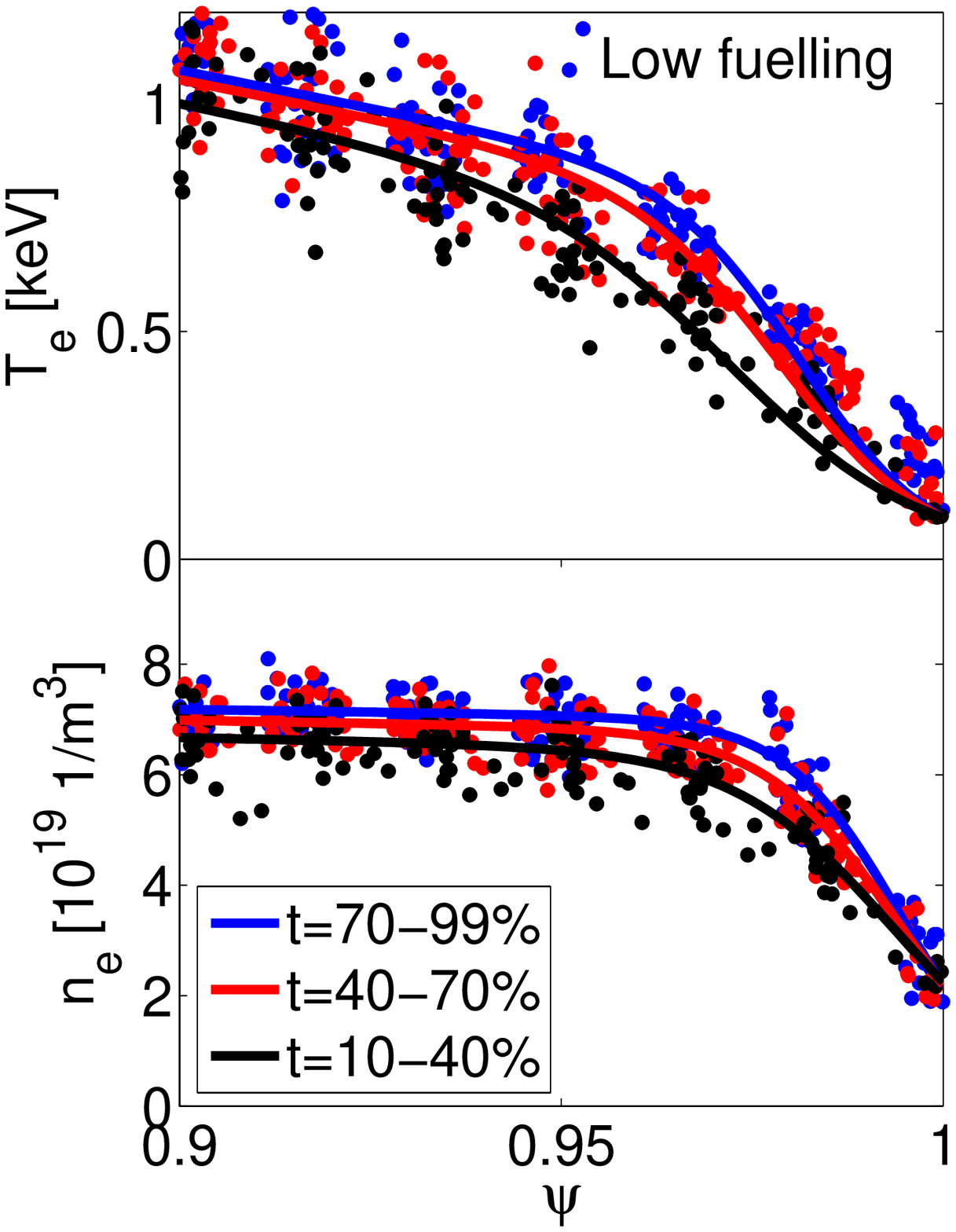,width=\linewidth}
\end{minipage}
\caption{The density and temperature profile evolution during the ELM cycle
 in high (left) and low (right) fuelling JET discharges. The labels
 represent the normalized time in the ELM cycle.}\label{profiles}
\end{figure}

\change{}{For this study, we have chosen two high triangularity ($\delta=0.41$) ELMy H-mode JET discharges
from a fuelling scan in the Carbon wall and divertor configuration with $I_p$=2.5 MA, 
$B_t$=2.7 T \cite{leyland}. The discharges were selected from opposite ends of the fuelling scan (but at
at sufficiently low density to avoid the transition to Type III ELMs), and 
with the highest available resolution of Thomson scattering data in the JET pedestal region.} The discharge
\#79503 has high fuelling ($2.6\times 10^{22}$e/s) and the discharge
\#79498 has low fuelling ($0.2\times 10^{22}$e/s). The density, temperature
and pressure profiles of the two plasmas during the ELM cycle are shown in
Fig. \ref{profiles}. The period right after the ELM crash (0-10\% of the
ELM cycle) is ignored because the data in the period is dominated by the
ELM crash and not very useful for analysis. Excluding it does not change
any of the conclusions presented in the paper. A more detailed discussion
about the data and fitting is given in \cite{leyland}. $Z_{\rm eff}$ is 2.0 for
the low fuelling case and 1.7 for the high fuelling case.
\added{ While both plasmas have large Type I ELMs, the high fuelling
case has also larger inter-ELM losses and the ELMs are classified as mixed
Type I/Type II: we focus our analysis on the Type I ELM cycle in both plasmas. }

\begin{figure}
\epsfig{file=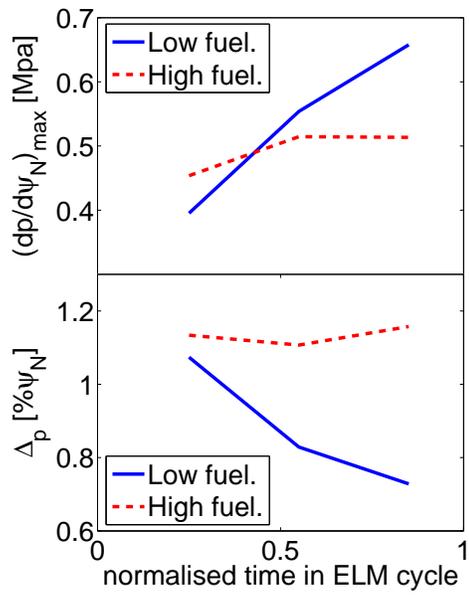,width=.5\linewidth}
\caption{The time evolution of the maximum of the pedestal pressure
 gradient and the pedestal width during the ELM cycle in the high and low
 fuelling JET discharges.}\label{dpwidth}
\end{figure}

As described in detail in \cite{leyland}, the pedestal behaviour between
ELMs is different in these two cases. In the low fuelling plasma, both the
temperature and density pedestal heights increase and the widths decrease
during the ELM cycle and the pressure gradient increases until the
ELM crash collapses the pedestal. \added{(Very similar profile evolution was observed
in the two other low fuelling discharges with the highest available resolution of HRTS.)} 
In the high fuelling case; the density pedestal height increases slightly
while the temperature profile stays unchanged after the quick recovery
following the ELM crash; initially (up to midway through the ELM cycle) the
maximum pressure gradient in the pedestal increases, but then saturates and
does not change for a large part of the ELM cycle. The width of the
pressure profile increases very slightly during the ELM cycle. The time
evolution of the pressure gradient and the pedestal width for both
discharges are shown in Fig. \ref{dpwidth}. 
\added{At the fully developed pedestal the ratios of the measured pedestal pressure
height and width (in normalised poloidal flux) to the EPED1 predictions 
are: 0.94 and 0.83 respectively in the low fuelling discharge (\#79498); 
and 1.38 and 1.35 respectively in the high fuelling discharge (\#79503).
The pedestal parameters in these particular discharges are entirely consistent 
with the trends over the fuelling scan that are illustrated in Figs.~16 and 17 of \cite{leyland}.}

\begin{figure}
\begin{minipage}{.5\linewidth}
\epsfig{file=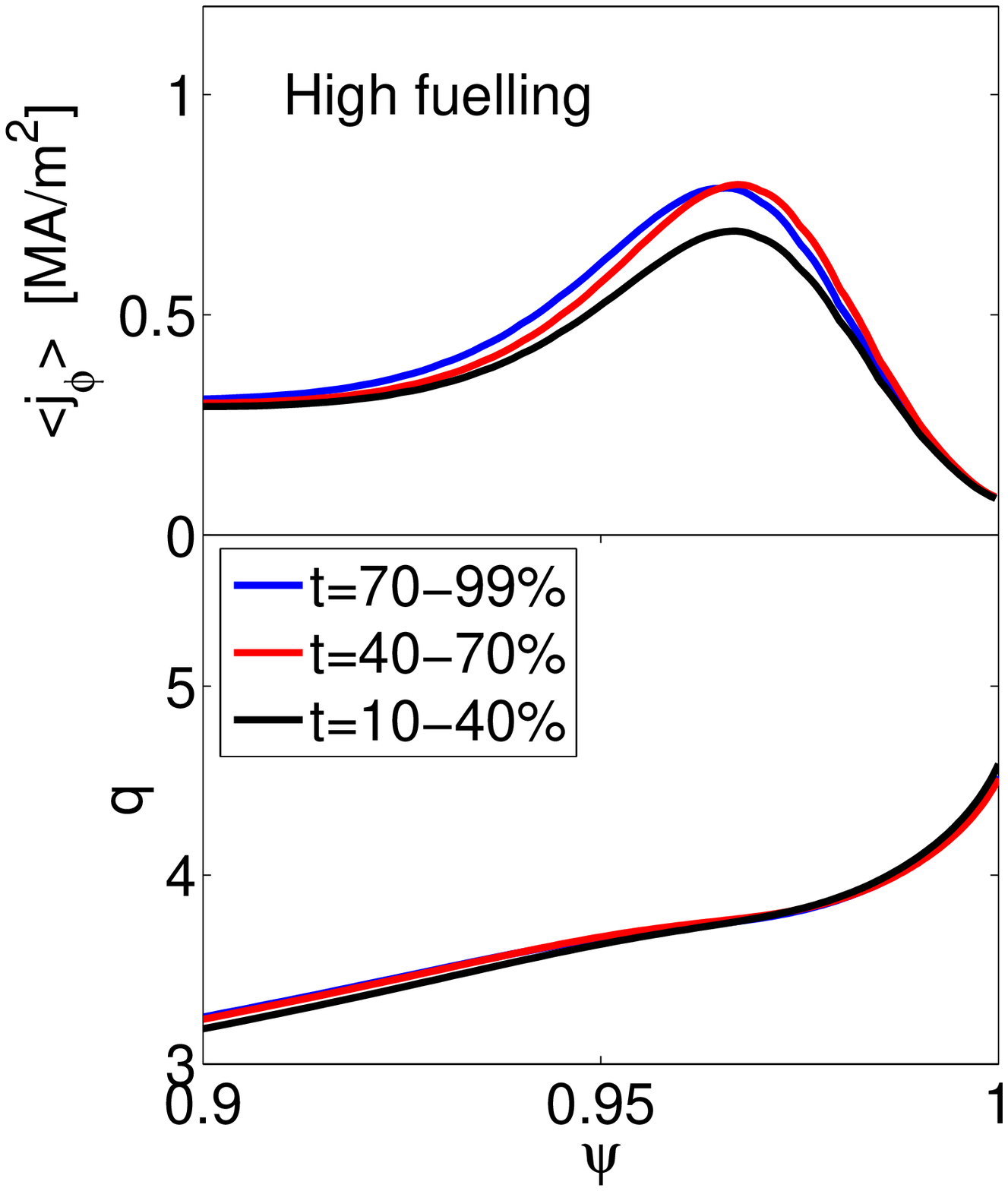,width=\linewidth}
\end{minipage}
\begin{minipage}{.5\linewidth}
\epsfig{file=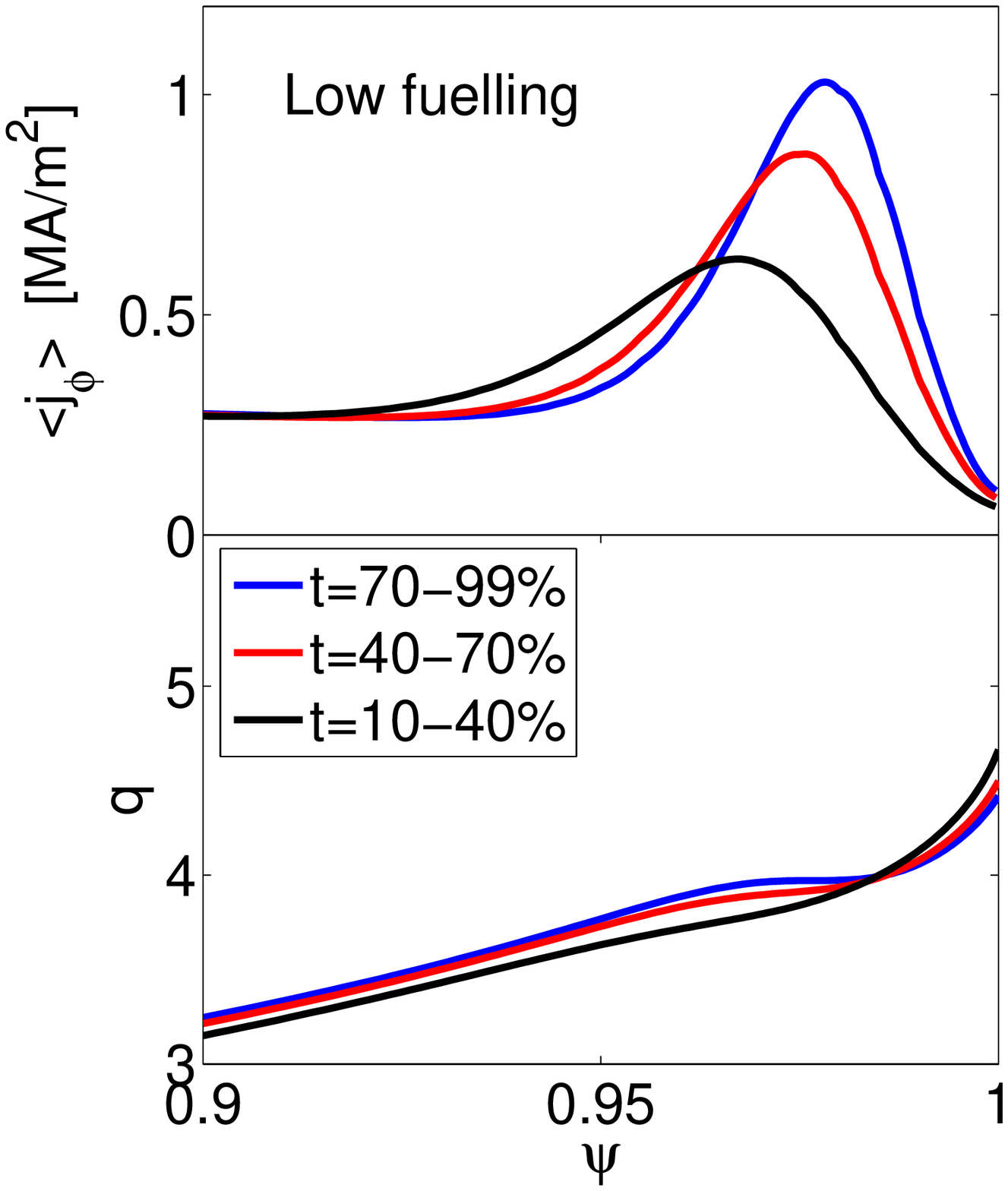,width=\linewidth}
\end{minipage}
\caption{The flux surface averaged toroidal current and q- profile
 evolution during the ELM cycle in high (left) and low (right)
 fuelling JET discharges. The labels represent the normalized time in the
 ELM cycle.}\label{jandq}
\end{figure}

In the equilibrium reconstruction the self-consistent bootstrap current is
assumed to follow the pressure profile without a delay. It has been shown
that this is a good approximation for current diffusion in similar plasmas
in ASDEX Upgrade \cite{burckhart}. The resulting toroidal current and
$q$-profiles are shown in Fig.\ref{jandq}. As can be seen, the $q$-profile
develops a flat part in the location of the maximum bootstrap current.

\section{MHD Stability analysis}
\subsection{Finite-$n$ MHD Stability Analysis}
The finite-$n$ MHD peeling-ballooning instabilities have been found to
limit the pedestal height in Type I ELMy H-mode plasmas in several devices
\cite{snydernf09}. In this paper, we use the MISHKA-1\cite{mishka}  ideal MHD code to investigate the finite-$n$ stability
of the plasma during the ELM cycle. We conduct two different types of scans
around the experimental equilibrium. First we map the stability boundaries
in $\alpha-<j_\phi>_{max}$-space. Here the normalized pressure gradient,
$\alpha$, is defined as \cite{alpha}
\begin{equation}
\alpha=\frac{-2\partial V/\partial
\psi}{(2\pi)^2}\left(\frac{V}{2\pi^2R_0}\right)^{1/2}\mu_0 \frac{\partial
p}{\partial\psi},
\end{equation}
where $V$ is the plasma volume, $R_0$ is the major radius and $p$ is the
pressure.  $<j_\phi>_{max}$ is the maximum of the flux surface averaged
toroidal current density in the pedestal region. In the scan we vary the
$dp/d\psi$ and $<j_\phi>$ profiles from the experimental profile using the
method described in detail in \cite{iternf}. This method allows the
profiles to be varied locally without introducing discontinuities.

\begin{figure}
\begin{minipage}{.5\linewidth}
\epsfig{file=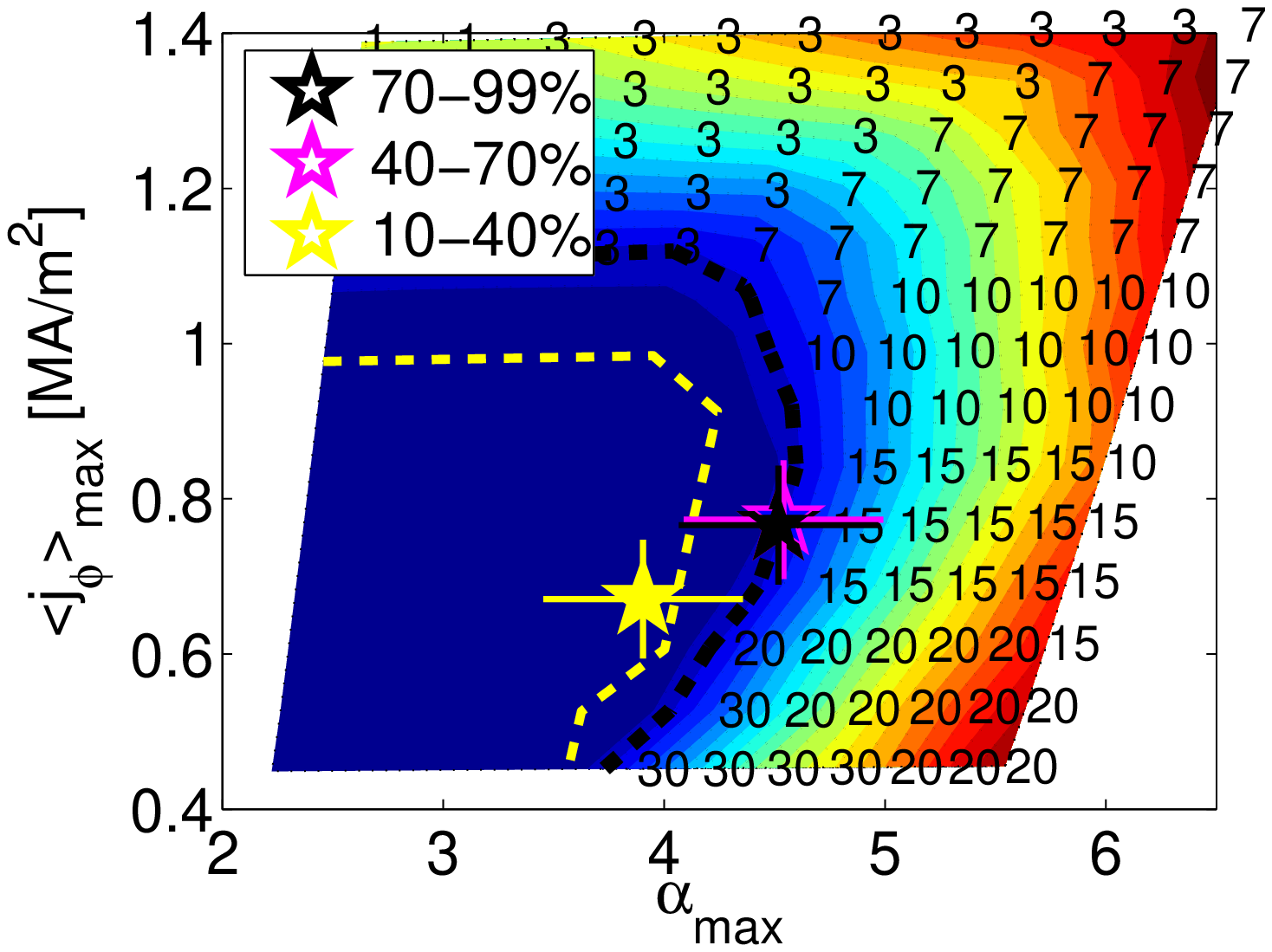,width=\linewidth}
\end{minipage}
\begin{minipage}{.5\linewidth}
\epsfig{file=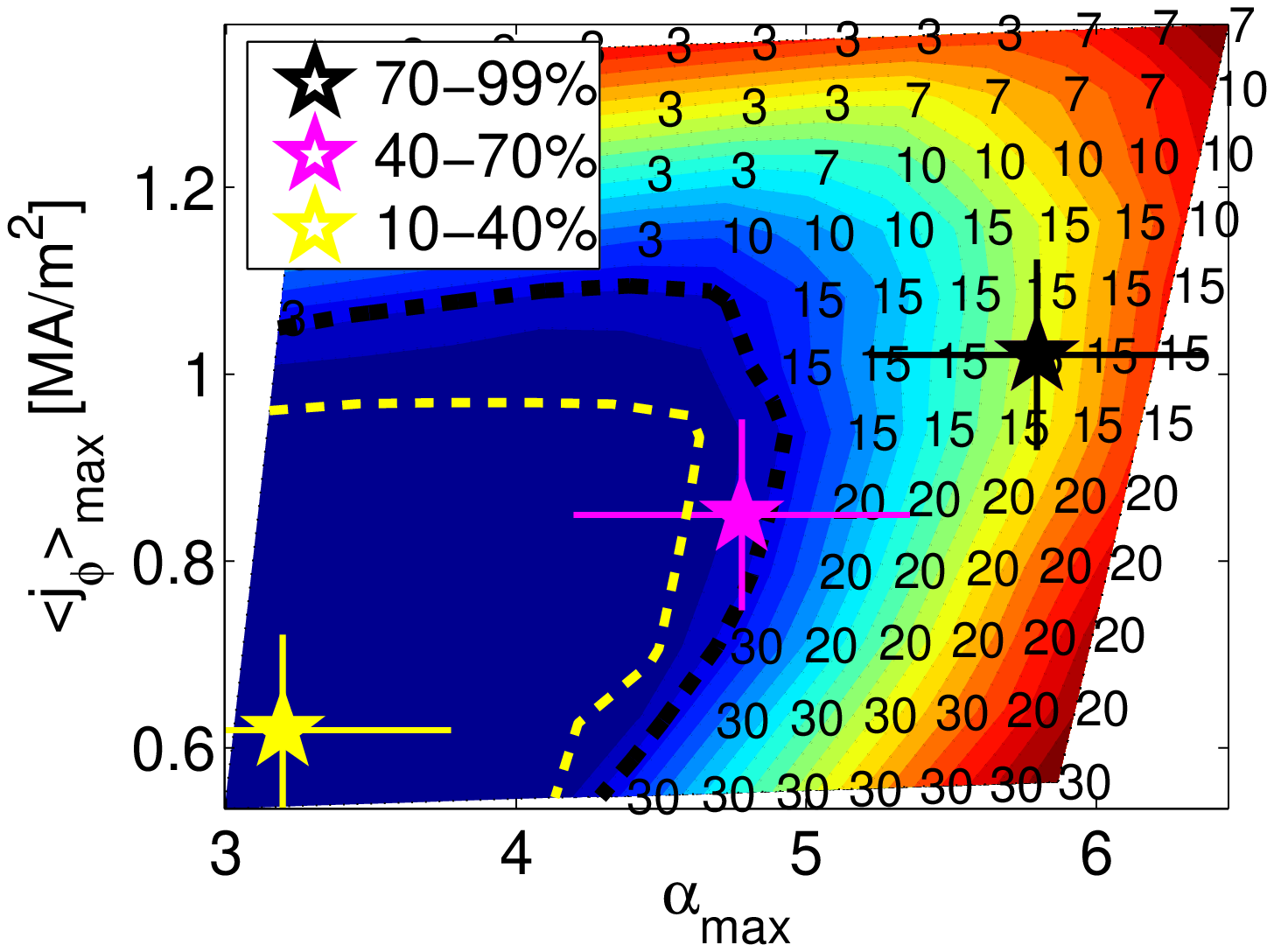,width=\linewidth}
\end{minipage}
\caption{The peeling-ballooning mode stability diagrams, 
\added{for toroidal mode numbers in the range $3<n<30$,} for the high
 fuelling (left) and low fuelling (right) plasmas during the ELM cycle. The
 color represents the growth rate of the fastest growing mode (red is
 stable) and the numbers give the toroidal mode number of the most unstable
 mode. The dashed lines shows the stability limit at $\gamma>\omega*/2$
 (black) and $\gamma>0$ (yellow). The stars with the cross show the position of the
 experimental equilibrium during the ELM cycle \added{ and their
 estimated errors \cite{leyland}. 
 }}\label{stabdiag}
\end{figure}

The finite-$n$ stability diagrams of the two discharges\added{, for toroidal mode numbers in the range $3<n<30$,} are shown in
Fig. \ref{stabdiag}. 
In the diagrams the color represents the growth rate,
$\gamma$, of the most unstable mode and the stability boundaries are drawn
for $\gamma=0$ and $\gamma=\omega_*/2$, where $\omega_*$ is half of the
maximum diamagnetic frequency for a given mode number in the pedestal. 
\added{The finite-$n$ modes that become unstable late in the ELM cycle are 
peeling-ballooning modes with significant drive from the edge current density gradient $J_{\parallel}^{\prime}$.
Such peeling-ballooning modes are global in nature, have growth rates that peak at finite $n$ 
and become stable as $n \rightarrow \infty$.}

In both cases the edge plasma starts the ELM cycle deep in the stable
region. In the low fuelling case the pressure gradient and the current
density keep increasing until the end of the ELM cycle, when the plasma has
crossed into the unstable region. 
Since the pedestal width of the last time
point is close to the width of the instrument function used in the fitting,
the deconvolution technique used in the fitting procedure may in this
particular case overestimate the gradient that would explain the last time
point being relatively deep in the unstable region. In the high fuelling
case the pressure gradient and the current density saturate long before the
ELM crash with the edge plasma close to marginal stability. As a conclusion
of the finite-$n$ stability analysis we can say that the result is
consistent with the assumption in the EPED model that the
peeling-ballooning modes are the ultimate limit for the pedestal height.

\subsection{$n=\infty$ MHD Stability Analysis}
\begin{figure}
\begin{minipage}{.5\linewidth}
\epsfig{file=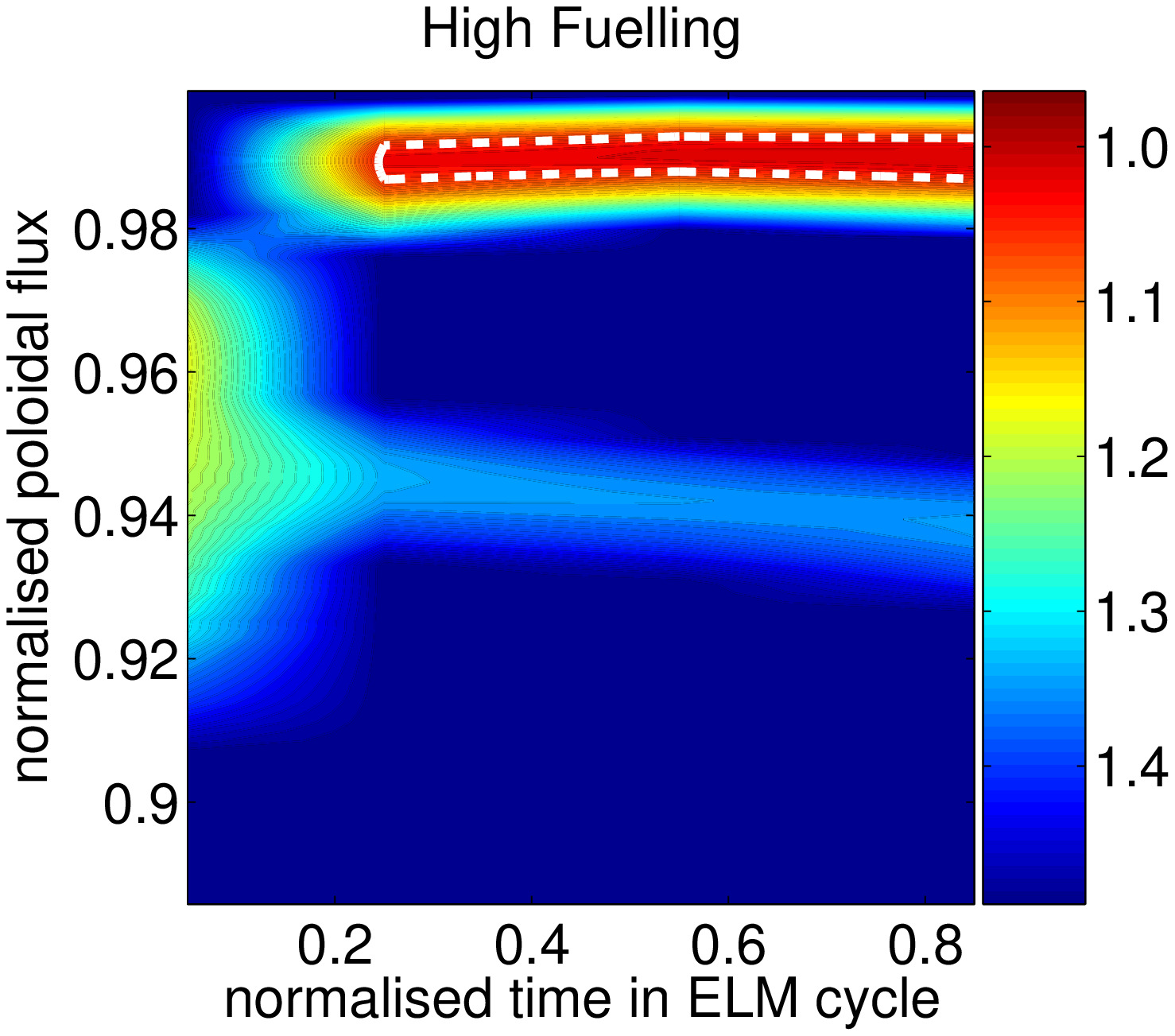,width=\linewidth}
\end{minipage}
\begin{minipage}{.5\linewidth}
\epsfig{file=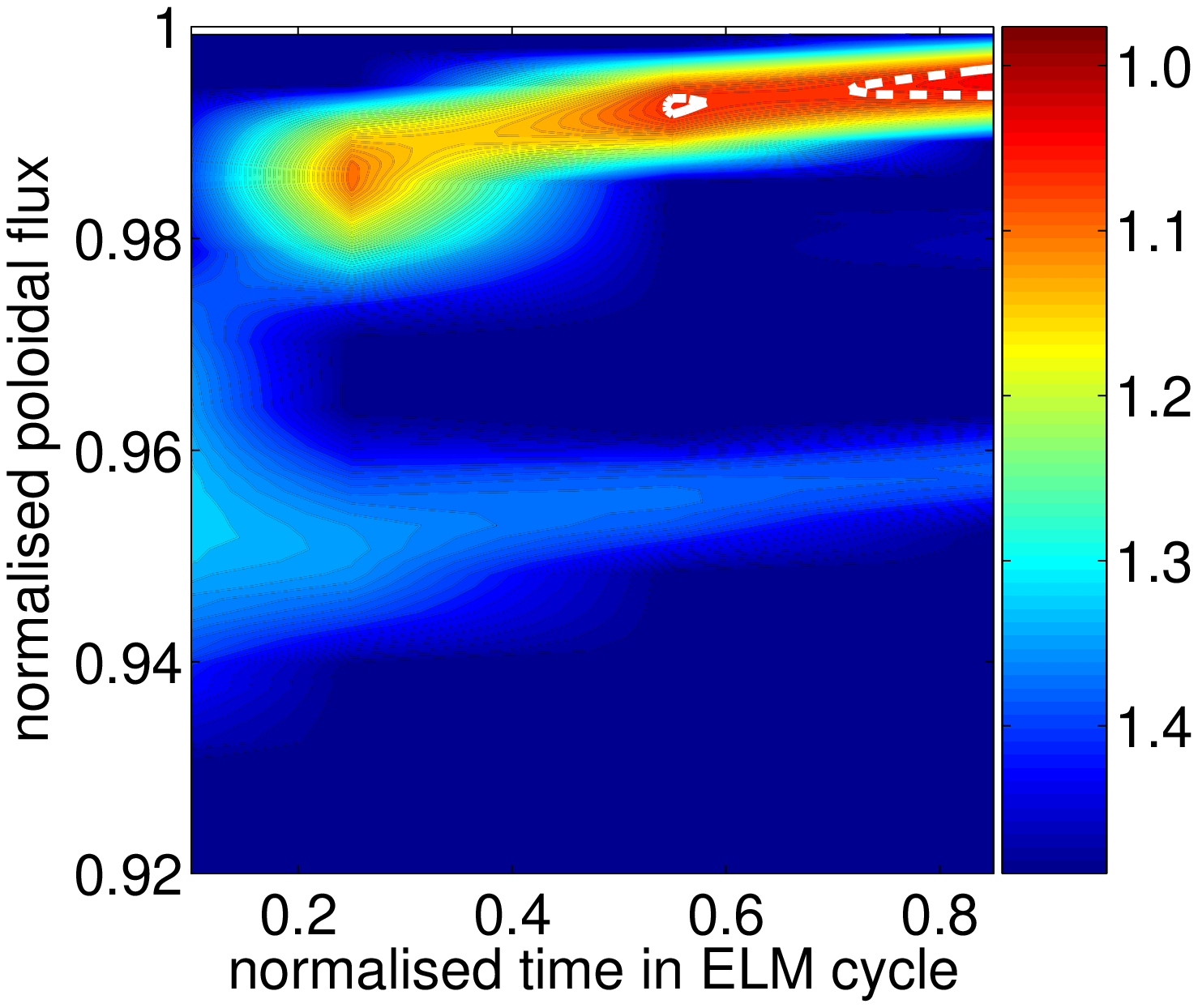,width=\linewidth}
\end{minipage}
\caption{The $n=\infty$ ballooning stability during the ELM cycle in the
 edge region for the high (left) and low (right) fuelling cases. The colors
 represent the \added{ value of $F=\alpha_{crit}/\alpha$, where $\alpha_{crit}$ is
the marginally ballooning stable value of $\alpha$}, i.e. regions below unity are unstable
 and above unity stable. The dashed line shows the region of marginal
 stability.}\label{ballooning}
\end{figure}

\change {It has been shown that \added{in the pedestal region of MAST plasmas} local
$n=\infty$ ballooning stability, which can be calculated very quickly for
an equilibrium, matches local kinetic ballooning mode (KBM) stability
calculated using a local gyro-kinetic code \cite{dickinsonppcf}.}
{It has been shown that $n=\infty$ MHD ballooning modes are unstable in the conditions 
of the MAST pedestal \cite{dickinsonppcf,dickinsonprl}, with stability boundaries that match closely those 
from local gyro-kinetic calculations of kinetic ballooning modes (KBMs).\footnote{Theoretically this is not surprising, as the 
kinetic result reduces to the MHD ballooning equation in the appropriate low frequency limit \cite{tangconnorhastie}.}} 
In the
analysis presented in \cite{dickinsonppcf} for the MAST plasma, the edge
plasma is relatively cold ($T_{e,ped}<200$ eV) resulting in a relatively
collisional plasma with a small bootstrap current peak. However, in the JET
discharges analyzed in this paper the pedestal temperature is much higher
($T_{e,ped}\approx 1$ keV) and the resulting bootstrap current peak is
large. As was shown in Fig. \ref{jandq} the bootstrap current peak flattens
the $q$-profile in the steepest pressure gradient 
\added{ region, locally stabilising ballooning modes.}
\added{ We investigate the $n=\infty$ ballooning
stability by scaling the normalised pressure gradient $\alpha$ locally,
checking the ballooning stability for each value of $\alpha$ and plotting
the value of $F=\alpha_{crit}/\alpha_{exp}$, where $\alpha_{crit}$ is the
marginally ballooning stable value of $\alpha$ and $\alpha_{exp}$ is the
experimental value of $\alpha$. Figure \ref{ballooning}
shows the evolution of $F$ during the ELM cycle in the two analyzed
JET pedestals. In the graph, the flux surfaces where $F>1$ are stable and
those with $F<1$ are unstable.} As can be seen, there is a narrow band of
unstable plasma between the steepest gradient and the edge. The steepest
pressure gradient region is stable through the ELM cycle,
\added{ as has been found previously in DIII-D and JET \cite{Miller,Lao,Lonnroth}.}
\added{This local stability is} due to the strong bootstrap current peak in the
pedestal. The``knee'' or the top of the pedestal is also stable, but closer
to the stability limit than the steepest gradient region.

\begin{figure}
\begin{minipage}{.5\linewidth}
\epsfig{file=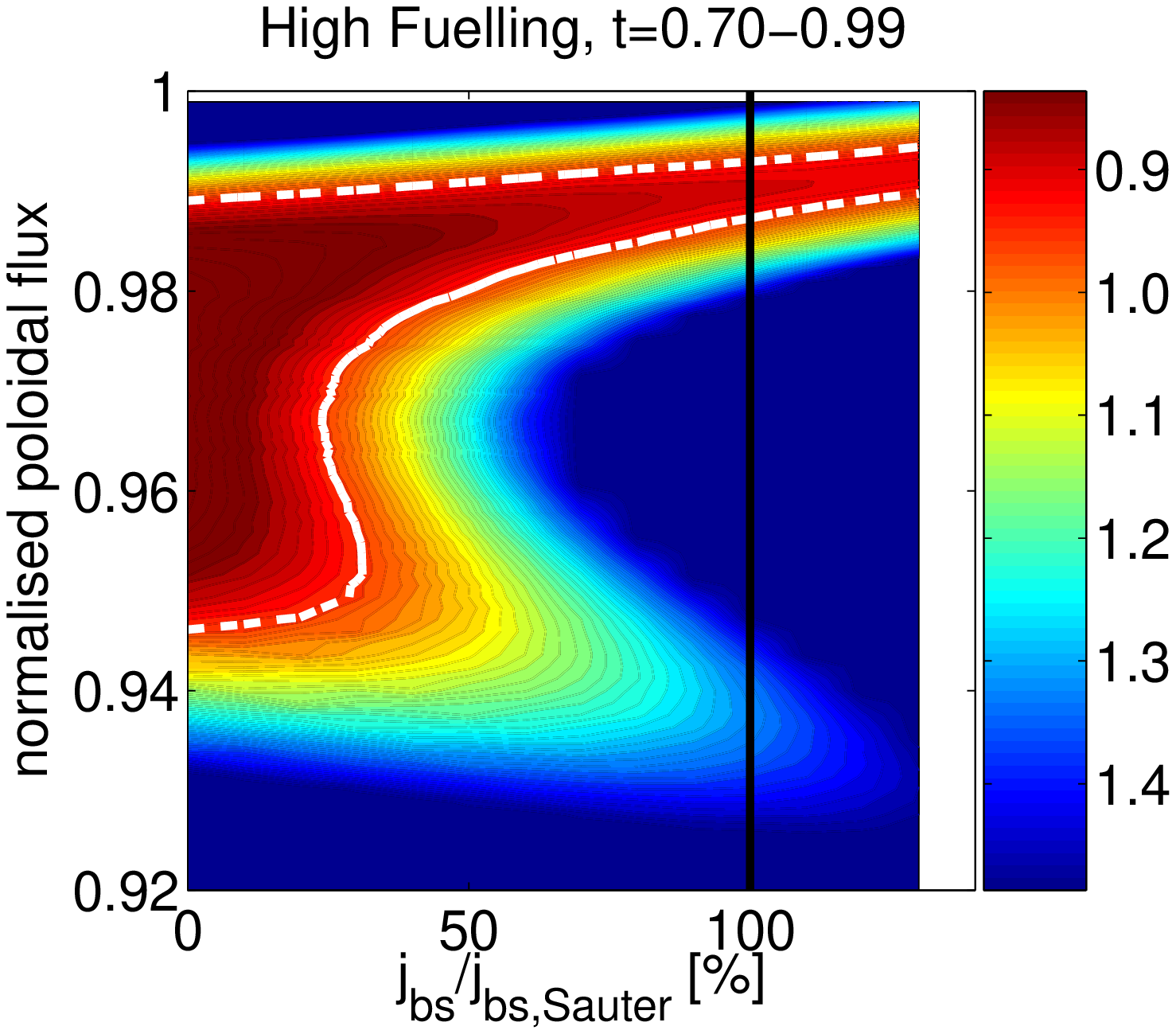,width=1.0\linewidth}
\end{minipage}
\begin{minipage}{.5\linewidth}
\epsfig{file=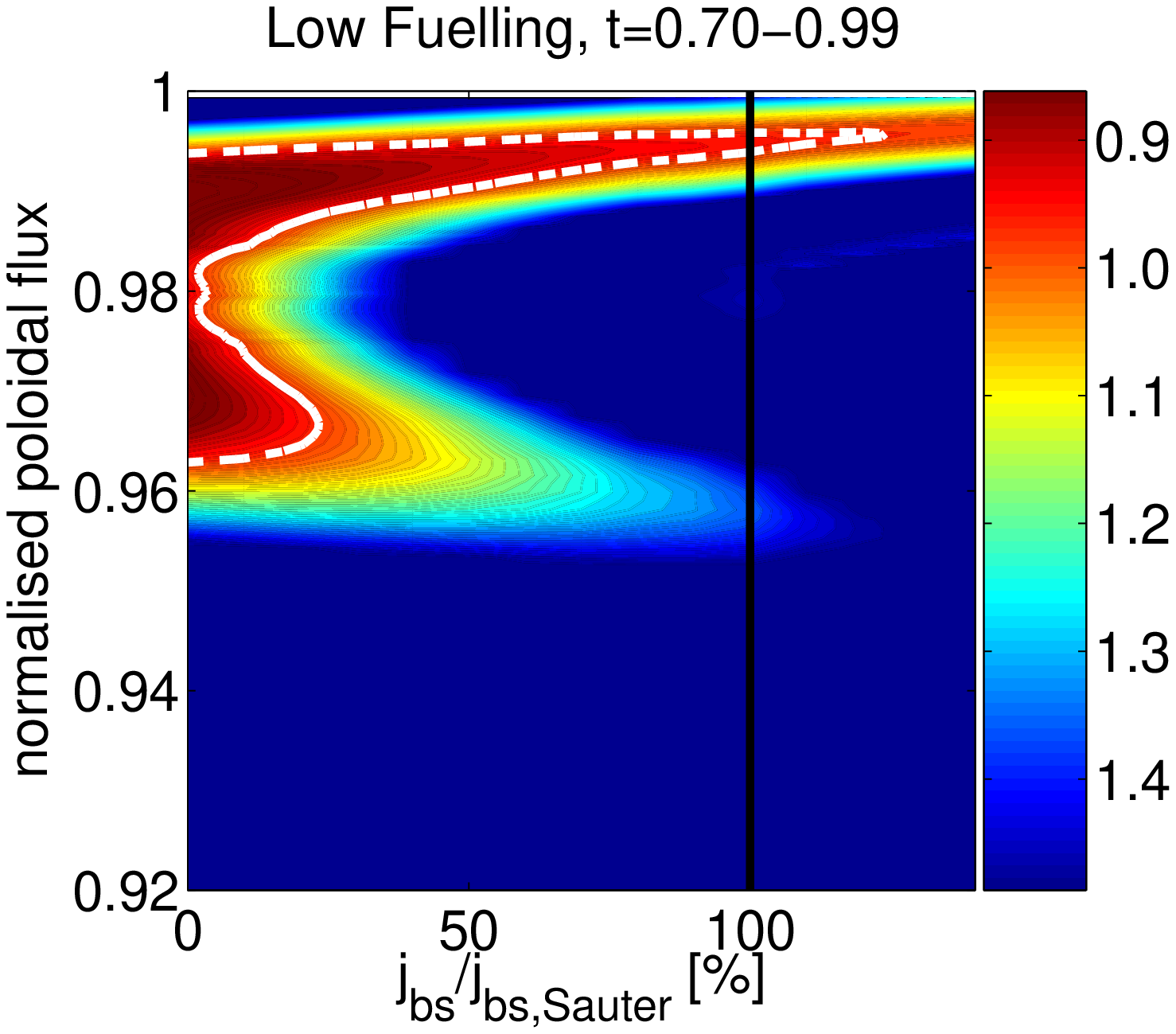,width=1.0\linewidth}
\end{minipage}
\caption{The $n=\infty$ ballooning stability as a function of bootstrap
 current taken into account in the equilibrium reconstruction for the
 pedestal region of high (left) and low (right) fuelling case at the end of
 the ELM cycle. The colors represent the stability (blue unstable, red
 stable) and the dashed \added{white} line shows the boundary of marginal
 stability. The vertical line represents the equilibrium with full
 bootstrap current given by the formula \cite{sauter}.}\label{bsscan}
\end{figure}

We demonstrate that the $n=\infty$ stability in the steepest pressure
gradient region is due to the bootstrap current peak by conducting a scan
in which we vary the amount of bootstrap current included in the
equilibrium reconstruction. The results shown in Fig. \ref{bsscan} are from
the equilibria at the end of the ELM cycle, but similar results are found
for other time points. As can be seen \added{more of} the pedestal region indeed becomes
unstable as the bootstrap current is reduced from the value given by the
formulas in Ref. \cite{sauter}. \added{Without the
large bootstrap current peak the entire steep pedestal region of the experimental
equilibrium becomes unstable to $n=\infty$ ballooning modes.}  As
can also be noted, the ``knee'' region becomes unstable with less reduction
of current than is required to destabilize the steepest pressure region.

\added{It is interesting to ask whether the surface with the steepest pressure 
gradient \footnote{Pressure gradients were obtained using $mtanh$-function fits.} 
and found to be locally ballooning stable due to the high bootstrap current, is  
unstable at an earlier time during the increase of the pressure pedestal. If ballooning instability 
arises earlier during the ELM recovery, KBMs might be expected to limit further local increases in pressure gradient. 
$dP/dr$ may still, however, increase in ballooning stable regions of the pedestal, until possibly becoming 
clamped by KBMs over a broader region.} 
\removed2{This is the idea behind the ballooning critical pedestal
(BCP) method used in the EPED 1.6 model \cite{eped16}.}
\added{If the steepest $dP/dr$ surface remains locally ballooning stable throughout the ELM recovery,
then the local analysis cannot predict any limit to the maximum $dP/dr$; if
KBMs are \addedd{also globally} stable, $dP/dr$ can grow until 
it becomes limited by another mechanism or by the ELM crash.}


\added{Fig. \ref{ballooning} demonstrated that the steepest pressure gradient
regions of the JET pedestals are locally stable to $n=\infty$ ballooning modes
throughout the ELM cycle. During the ELM cycle, however, both the
width of the pedestal and the location of the steepest pressure gradient evolve. 
In order to study in isolation the effect of pedestal pressure profile steepening, 
we conduct an artificial equilibrium scan based on the profiles from the last point of the ELM cycle.
In this scan we have varied the height of the temperature pedestal, obtained 
self-consistent equilibria with modified bootstrap current profiles, and tested 
$n=\infty$ ideal ballooning stability.}
\added{The ideal $n=\infty$ ballooning unstable region is plotted as a function of $P_{\rm ped}$ in Fig. 7.
We note that no unstable flux surfaces were found for pedestal pressure values lower than those shown 
in the plot. In both high and low fuelling plasmas the ballooning unstable region moves slightly inwards as 
$P_{\rm ped}$ is reduced, but the maximum $dP/dr$ surface in the middle of the pedestal remains locally ballooning 
stable throughout both scans. At high fuelling the ballooning unstable region reaches into the outer part of the 
central-half of the pedestal at lower $P_{\rm ped}$, but at low fuelling the unstable region remains in a 
narrow region closer to the plasma edge.
Only the very narrow outer region (especially in the case of the low fuelling plasma) 
appears to be limited by a local high $n$ ballooning instability.
We have also performed, over these $P_{\rm ped}$ scans for low and high 
fuelling plasmas, global ideal MHD stability analysis at high finite-$n$ 
($35<n<70$, extending the range of Sec. 3.1). These high $n$ global modes were found to be stable throughout both scans. 
Fig. \ref{stabdiag} shows that peeling-ballooning modes at lower $n$  become unstable as $P_{\rm ped}$ approaches 
its pre-ELM value.}

\section{Gyro-kinetic analysis}

Finite-$n$ MHD stability of the plasma edge provides hard limits
that 
\added{cannot be breached without} the loss of a considerable
part of the plasma as happens during an ELM. However, more benign
turbulence driven by local micro-instabilities can also limit the edge
gradients through transport. 
\added{KBMs are micro-instabilities that trigger stiff transport above a critical pressure gradient,
potentially limiting the pressure gradient to this critical value. Other types of mode may also be 
unstable in the pedestal, triggered by density or temperature gradients. Such modes also cause transport 
that affects the evolution of the pedestal, but do not directly limit the pressure gradient 
to a specific critical value.}

To study the micro-instabilities in the pedestal region during the ELM
cycle in more detail, we use local gyro-kinetic analysis, and we consider
modes with size of the order of ion Larmor radius, which is generally small
compared to the equilibrium scale lengths. At the top of the pedestal this
requirement is easily met ($L_T/\rho_i\approx 80$ and $L_n/\rho_i\approx
800$), but in the steep gradient region the gyrokinetic expansion parameter
is larger (both $L_n/\rho_i$ and $L_T/\rho_i$ $\approx
10-20$). 
\added{We note that recent global gyro-kinetic analysis, agreeing with local flux-tube simulations at high $n$ ($n>21$), 
has demonstrated the existence of near-threshold conditions for kinetic ballooning modes in D-IIID H-mode pedestals
\cite{wan}.}

\subsection{Gyrokinetic stability in the steep gradient region}
In the previous section we found that 
\added{the JET plasmas are locally stable to $n=\infty$ ideal MHD ballooning modes across most of the pedestal, apart from in a narrow outer region.}
It is well known \removed{for long} that \removed{the} ideal MHD ballooning modes and kinetic
ballooning modes have very similar drive mechanisms
\cite{kotchenreutherpf}.  
\added{Ideal MHD $n=\infty$ stability in the MAST pedestal has been found to be a 
reliable indicator of local
kinetic ballooning mode (KBM) stability calculated by a local gyro-kinetic
code \cite{dickinsonppcf}. We will now test if this is true also for the JET
pedestal with higher bootstrap current.}
For this analysis we use the local electro-magnetic \added{gyrokinetic} code GS2
\cite{gs2}. Both the ions and electrons are treated kinetically and
the collisions are taken into account. The effect of sheared flow, while
important for modes with growth rates similar to the shearing rate, may be
less important for KBMs because once the plasmas crosses the stability
boundary for KBMs, the growth rate increases very rapidly
\cite{snyder01,roach05} resulting only in a small increase in the critical
pressure gradient. Therefore, the flow shear is neglected in this
analysis. Future studies are planned to include the effect of the flow
shear.

The linear \added{ local} gyro-kinetic stability analysis at
$k_\theta\rho_i<$0.2 ($k_\theta$ is the poloidal wave number and $\rho_i$
is the ion Larmor radius) finds no kinetic ballooning modes (KBM) in the
JET pedestals. This agrees with the ideal MHD $n=\infty$ stability
described earlier. At higher 
\added{$k_\theta \rho_i$} ($0.2<k_\theta\rho_i< 2$) we find
ion temperature gradient/trapped electron modes (ITG/TEM) and at even higher 
\added{$k_\theta \rho_i$ ($k_\theta \rho_i>5$)}  
electron temperature gradient (ETG) modes. 
\added{These modes have growth rates that do not increase with $\beta$
at constant $R/L_p$
(where $\beta$ is the ratio of plasma pressure to magnetic energy). ITG and ETG 
modes} are
not destabilized by increasing density gradient making them unlikely
candidates for limiting the pedestal density gradient. Therefore, while it
is possible that these modes play a role in the limiting of the pedestal
temperature gradient and slowing down the pedestal recovery, in this paper
we concentrate on KBMs that are destabilized by increasing pressure
gradient and have very high growth rates producing 
\added{stiff} turbulent
transport \added{in all channels} once the stability limit is exceeded \cite{rewoldt}.

We scan the plasma $\beta$ \added{locally} in GS2 (\change{i.e. not}{without} solving for a new
self-consistent equilibrium using HELENA) to find the \added{local} stability limit for
KBMs\added{, which are found to be unstable at $k_\theta\rho_i\sim 0.1-0.2$ in these JET pedestals. 
The equilibrium scale length exceeds the corresponding rational surface spacing 
close to the pedestal top, but this local approximation is
challenged in the steepest gradient region.} As can be seen in 
Fig. \ref{kbm_beta}, the amount by which $\beta$
must be increased in order to destabilize KBMs matches very well with the
$F$ parameter in the $n=\infty$ ideal MHD analysis throughout the pedestal
region. So, while the validity of the gyrokinetic
theory in the pedestal region is marginal due to small $L_n/\rho_i$ and
$L_T/\rho_i$ values, the \change{numerical results for}{local stability of} KBMs 
still agrees very well
with ideal MHD \change{results}{stability at $n=\infty$}. We can also see that the regions closest to the
stability limit are the bottom of the pedestal ($\psi\approx 0.99$) and the
``knee'' of the pedestal ($\psi\approx 0.94$ in high fuelling case and
$\psi\approx 0.96$ in the low fuelling case). The steepest part of the
pedestal is \added{locally} very stable \change{for}{to} KBMs.

The reason for the good \change{stability for}{local stability to} KBMs in the pedestal is the large
bootstrap current that lowers the flux-surface averaged magnetic shear in
the steep pressure gradient region. As was shown in the previous section,
reducing the bootstrap current that \added{ is} included in the equilibrium
reconstruction destabilizes the local $n=\infty$ ballooning modes. Exactly
the same happens with the KBMs in a \added{ local} gyro-kinetic simulation. Figure
\ref{kbm_bs} shows the \added{local} KBM growth rate in the pedestal region when no
bootstrap current has been taken into account. In the low fuelling case,
the \added{local} KBM growth rate has two peaks: one between the edge and the maximum
pressure gradient and the other between the top of the pedestal and the
maximum pressure gradient. These peaks match very well with the ideal MHD
$n=\infty$ ballooning unstable regions, i.e. the most $n=\infty$
ballooning unstable regions have also the highest \added{local} growth rates for KBMs. In
the high fuelling case, the \added{local} KBM growth rate peaks just outside the steepest
gradient region. Also in this case the \change{KBM unstable region}{region locally unstable to KBMs} matches
very well with the region unstable to ideal \added{$n=\infty$} MHD ballooning modes. 

\subsection{Gyrokinetic stability at the pedestal top}
In a gyro-kinetic analysis of the MAST tokamak, it was found that the
plateau region at the top of the pedestal is unstable to micro-tearing
modes (MTM) \cite{dickinsonppcf1,dickinsonprl,cmriaea}. The identification
of the micro-tearing modes in the JET pedestal top is made more difficult
by other unstable micro-instabilities. In the JET pedestal top, ITG modes
are the dominant instability. In order to identify possible sub-dominant
micro-tearing modes, we make the equilibrium up-down symmetric and in GS2
suppress even parity modes such as the ITG. Changing the plasma shape to
up-down symmetric has very little effect on the growth rates of the
dominant mode. When the ITG modes are suppressed, we find sub-dominant odd
parity micro-tearing modes. Figure \ref{fig:mutearitg} shows the growth
rates of the ITG (unsuppressed case) and MTM (when ITG is suppressed) at
the ``knee'' of the pedestal for both cases at the end of the ELM
cycle. The growth rate spectra for different time points is qualitatively
similar.

Similar to what was found in MAST \cite{dickinsonprl,dickinsonppcf1} and
unlike micro-tearing modes found in the core, the mode structure of the
pedestal top micro-tearing mode is not very extended along the field lines
as shown in Fig. \ref{fig:muteareigenfunc}. We also find that this mode is
relatively insensitive to the collisionality (growth rate increases
slightly with decreasing collisionality) and is unstable even when
collisions are 
\added{dropped from} the gyro-kinetic simulation. The growth rate of
the MTM at the pedestal top decreases with increasing density gradient and
increases with increasing temperature gradient as in
\cite{dickinsonprl,dickinsonppcf1,cmriaea}. 
\change{The MTM stabilization by the
density gradient means that MTMs can contribute to the transport of the
pedestal top region until they are suppressed by the increasing edge
density gradient as described in \cite{dickinsonprl}.}
{MTMs can contribute to the transport of the
pedestal top region, but would become suppressed by an increasing edge
density gradient as described in \cite{dickinsonprl}.}

\section{Conclusions}
\added{This paper presents finite-$n$ MHD stability analyses of carefully reconstructed 
experimental pedestal equilibria from JET H-mode plasmas with Type I ELMs, with both
high and low gas fuelling. At the end of the ELM cycle these pedestals are, 
within experimental uncertainties, limited by finite-$n$ peeling-ballooning modes. 
This confirms previous results from JET and various other devices, and is
consistent with the EPED model assumption that peeling-ballooning modes
limit the pedestal. Earlier in the ELM cycle both plasmas are shown to be
stable to finite-$n$ peeling-ballooning modes over the range $3<n<30$. In the high fuelling case the pedestal 
saturates close to the peeling-ballooning stability boundary around half-way
through the ELM cycle, while in the low fuelling case the pedestal pressure 
gradient keeps increasing until the ELM crash.}

In the pedestal region, the stability boundaries obtained using local ideal
MHD $n=\infty$ ballooning mode analysis and local linear gyro-kinetic
analysis for kinetic ballooning modes show very good correspondence. This
is consistent with MHD $n=\infty$ ballooning mode stability providing a
reasonable 
\added{indicator} for the \added{local} stability of kinetic
ballooning modes. 
\added{Most of the pedestal region in these discharges is found to be locally stable to 
both modes, due to reduced magnetic shear caused by high bootstrap current 
in the steep pressure gradient region. 
The steepest pressure gradient surface of the pedestal remains locally stable throughout artificial scans about the pre-ELM equilibria, 
where the bootstrap current is computed self-consistently as the pedestal height is varied from a low value to its pre-ELM value.
Furthermore, global MHD calculations find that all toroidal mode numbers in the range $35<n<70$ (i.e. at higher finite-$n$ than 
peeling-ballooning modes) are stable throughout these scans. 

We note that our local ballooning and finite-$n$ peeling-ballooning stability
results could be sensitive to any significant errors 
in the bootstrap current formula by Sauter {\em et al} \cite{sauter} or in other equilibrium parameters. 
The sensitivity of local KBM stability to the bootstrap current density points to the need for accurate measurement of the edge
current profile to validate the neoclassical models for the bootstrap current.
Large bootstrap current density and low magnetic shear are expected in ITER pedestals,
which may exclude KBMs from limiting the pedestal pressure gradient \cite{ssiter}.}

\added{ In the low fuelling JET plasma, the observation that the maximum 
pedestal pressure gradient increases over the ELM cycle, 
combines with the results from $n=\infty$ MHD and local gyro-kinetic analysis, 
to suggest that the maximum pedestal pressure gradient is not limited by kinetic 
ballooning modes during this ELM cycle. 
Furthermore, the narrowing of the pedestal through this ELM cycle is 
very different to typical ELM cycles in DIII-D and MAST where the
pedestal broadens with a constant maximum pressure gradient
\cite{groebner,dickinsonppcf}.}

\added{These JET experiments were} found to be consistent with the pedestal being
ultimately limited by peeling-ballooning modes. 
\added{The peeling-ballooning mode limit for the pedestal top pressure is relatively
insensitive to the pedestal width \cite{eped16,ssiter}, and}
it may not be necessary to get the
width exactly right in order to have a reasonable prediction for the
pedestal height.

Gyro-kinetic analysis also found micro-tearing modes at the JET pedestal
top with similar features to those found earlier in the MAST pedestal
top \cite{dickinsonppcf1,cmriaea}. 
\change{However, in JET the micro-tearing modes are sub-dominant to ITG modes.}
{In these JET plasmas the peak growth rate of the microtearing modes is approximately
half the peak growth rate of ITG modes. 
MTMs have been shown to be important, even when they are stable, in nonlinear electromagnetic 
simulations of ITG turbulence \cite{hatch}, and could play a role at the JET pedestal top
where they are unstable but sub-dominant to ITG modes.}

\added{Our analysis finds that in these JET plasmas KBMs are locally stable in most of the pedestal 
(inboard of and including the maximum $dP/dr$ surface) 
and locally unstable in a narrow region at the extreme edge. 
These local results suggest that KBMs have little impact on the evolution of most of the pedestal in 
these discharges, though we note that sheared flows and 3D equilibrium effects have not been included in this analysis.
Gyrokinetic analysis has limitations in the pedestal when the ion Larmor radius 
approaches the equilibrium scale length, and the accuracy of local gyrokinetics is further restricted 
in the steep pedestal where the rational surface spacing approaches the equilibrium scale length. The latter limitation can be 
alleviated through global gyrokinetic calculations that include radial equilibrium variation.
\addedd{Our local analysis cannot exclude the possibility that a globally
unstable KBM plays a role in the pedestal, but higher finite-$n$ MHD
ballooning mode calculations find no instability over the
range $35 < n < 70$.}
Investigation of this possibility requires a global gyrokinetic calculation, which 
will be considered in future work.}

\section*{Acknowledgements}
\added{The authors are grateful to Jack Connor, Jim Hastie, 
Phil Snyder and Howard Wilson for several stimulating discussions.} 
This work was partly funded by the RCUK Energy Programme under grant
EP/I501045 and the European Communities under the contract of Association
between EURATOM and CCFE and carried out within the framework of the
European Fusion Development Agreement. The views and opinions expressed
herein do not necessarily reflect those of the European Commission. The
gyrokinetic calculations were carried out on HECToR supercomputer (EPSRC grant
EP/H002081/1).

\pagebreak{}
\listoffigures

\begin{figure}
\begin{minipage}{.5\linewidth}
\epsfig{file=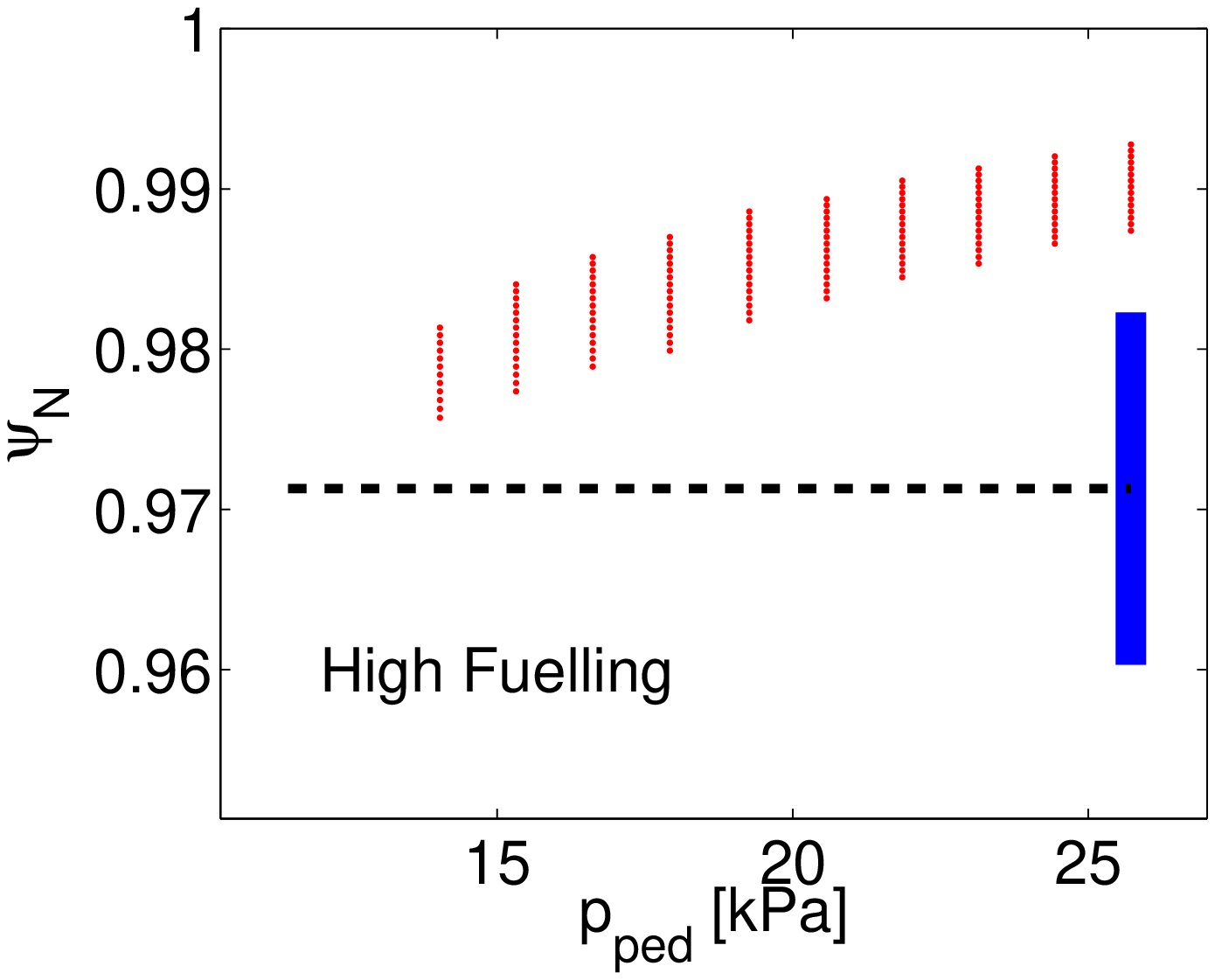,width=1.0\linewidth}
\end{minipage}
\begin{minipage}{.5\linewidth}
\epsfig{file=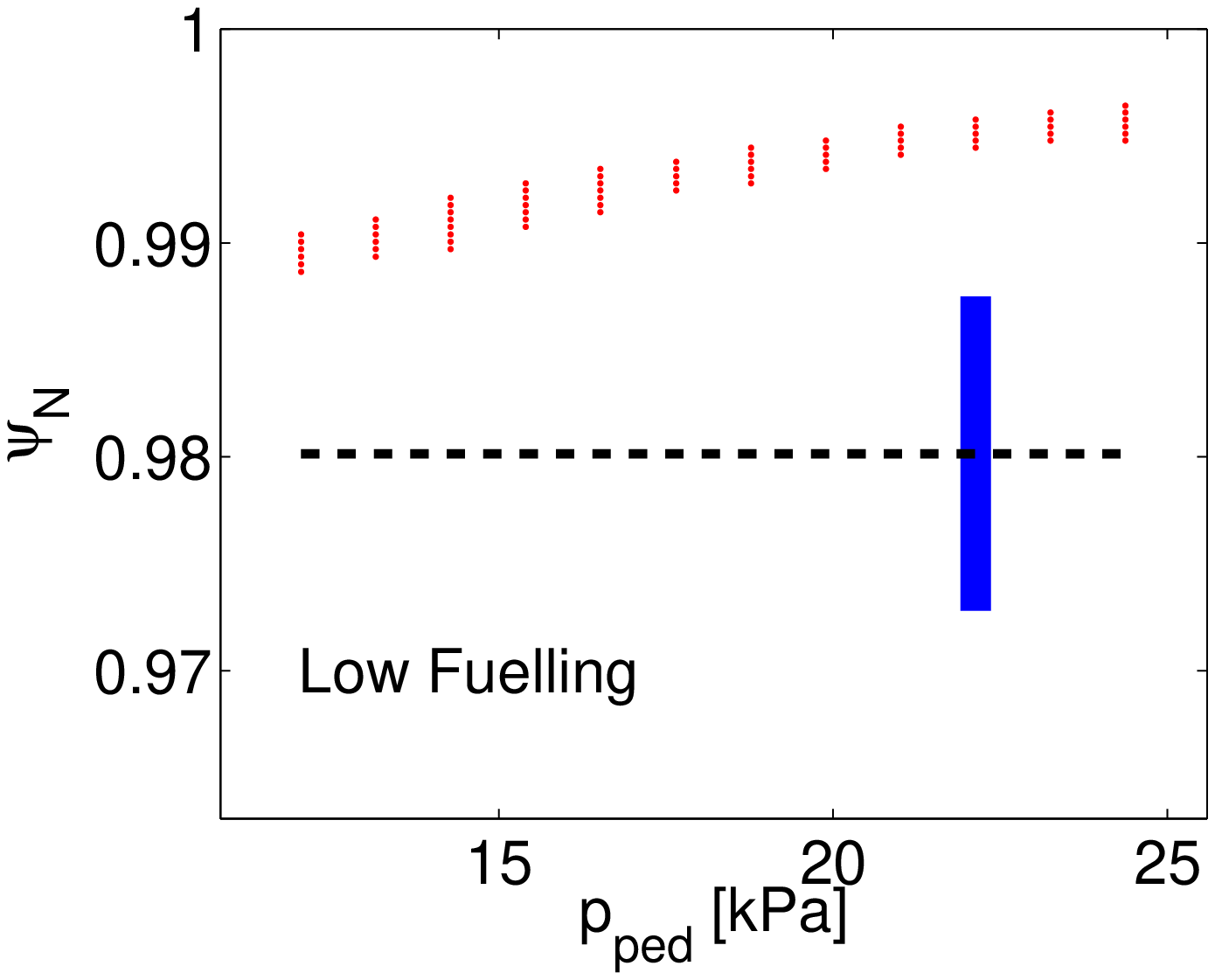,width=1.0\linewidth}
\end{minipage}
\caption{The $n=\infty$ ballooning unstable region in high (left) and
low fuelling (right) cases when the pedestal top temperature is varied
self-consistently. The red dots show the ballooning unstable flux surfaces.
The blue line represents the region of the pressure pedestal from 25\% to
75\% of the height and is placed at the value of the experimental pedestal
height. The black dashed line shows the steepest gradient
location.}\label{fig:tescan}
\end{figure}

\begin{figure}
\begin{minipage}{.5\linewidth}
\epsfig{file=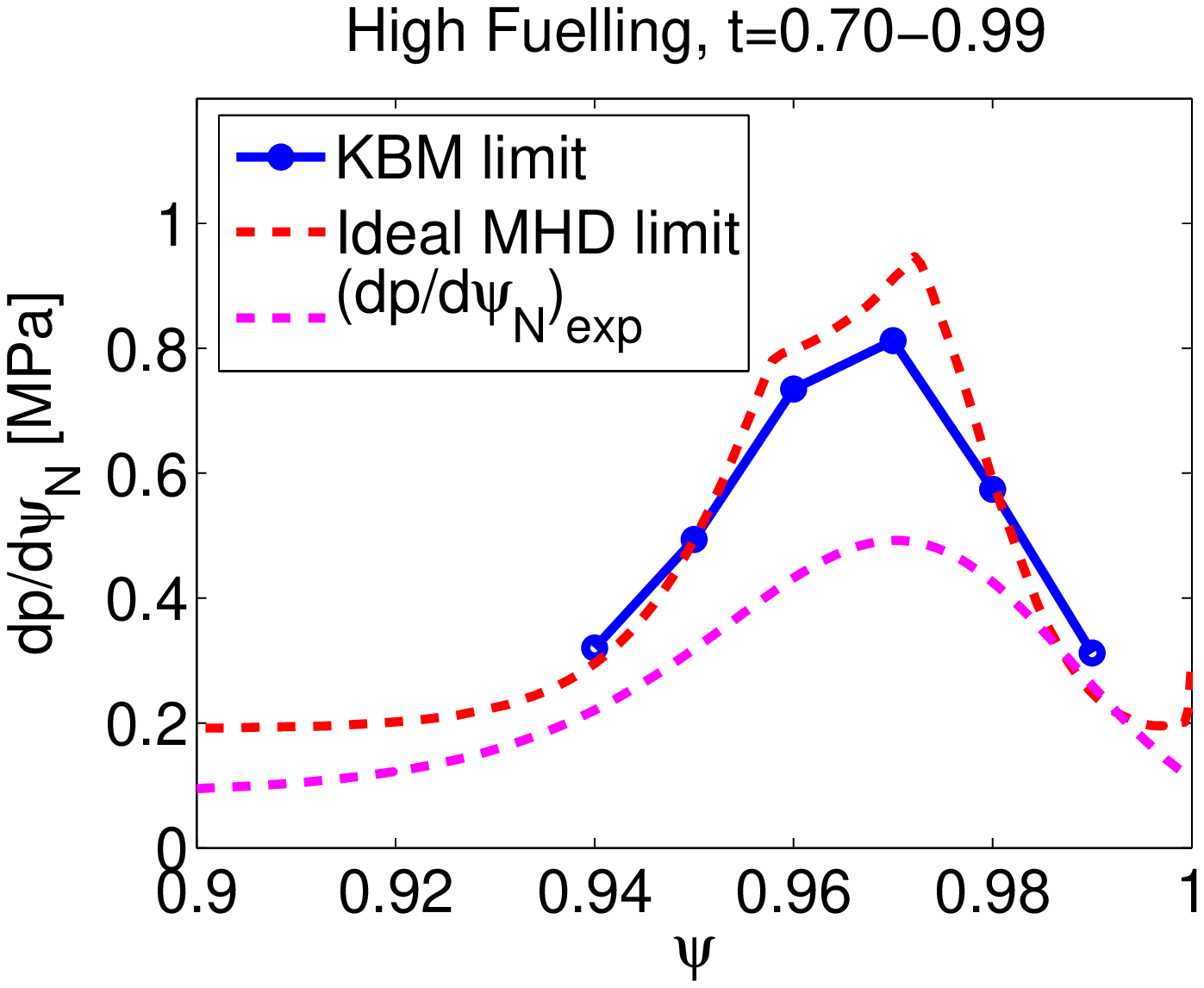,width=\linewidth}
\end{minipage}
\begin{minipage}{.5\linewidth}
\epsfig{file=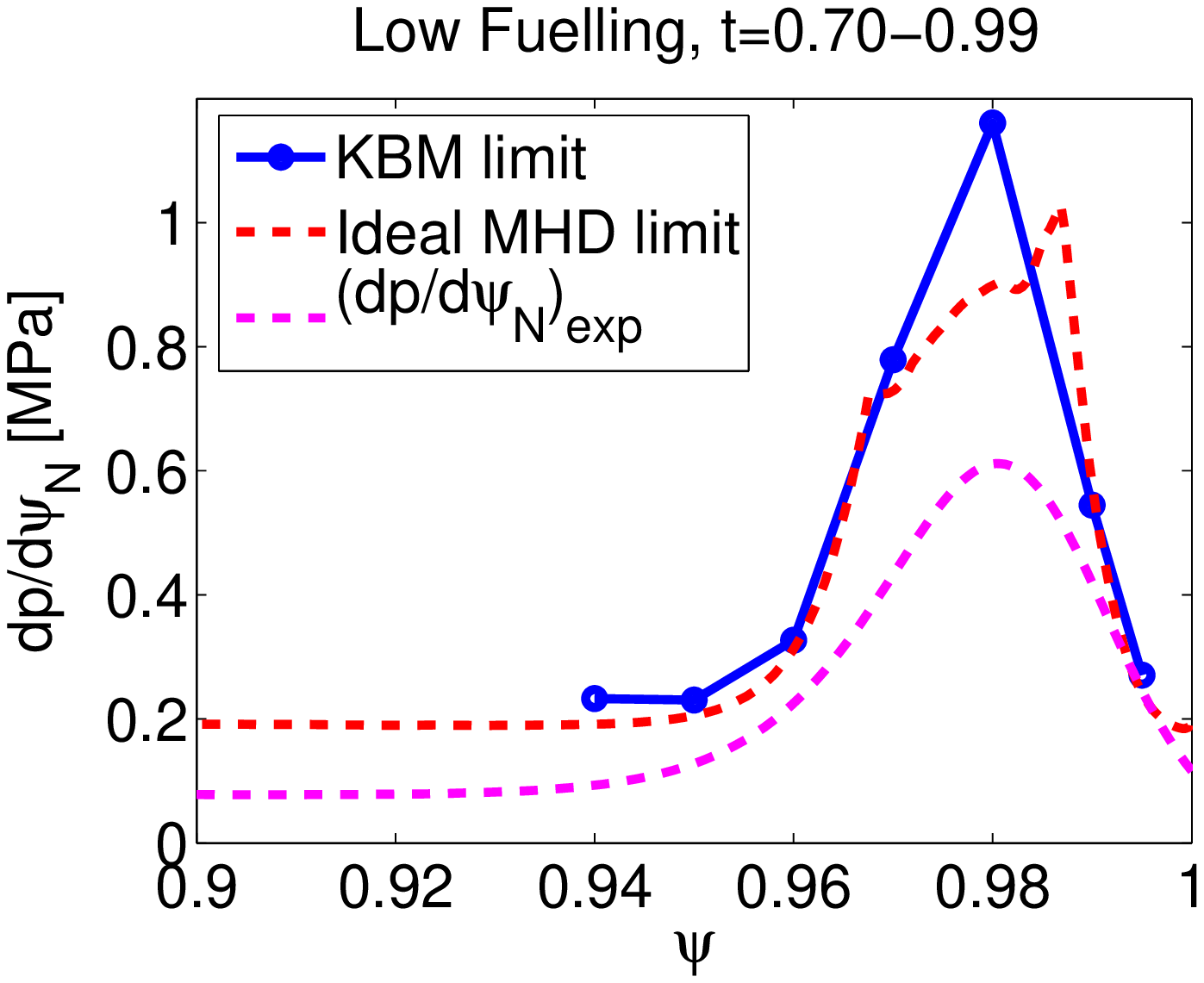,width=\linewidth}
\end{minipage}
\caption{The gyro-kinetic stability limit of KBMs (solid blue line) in high (left) and low
 (right) fuelling pedestals at the end of the ELM cycle as a function of
 poloidal flux. For comparison, the $n=\infty$ ideal MHD ballooning
 stability limit (dashed red line) and the experimental pressure gradient
 (dotted magenta line) are also shown. The stability limits represent the
 value the pressure gradient can be increased before becoming
 unstable}\label{kbm_beta}
\end{figure}

\begin{figure}
\begin{minipage}{.5\linewidth}
\epsfig{file=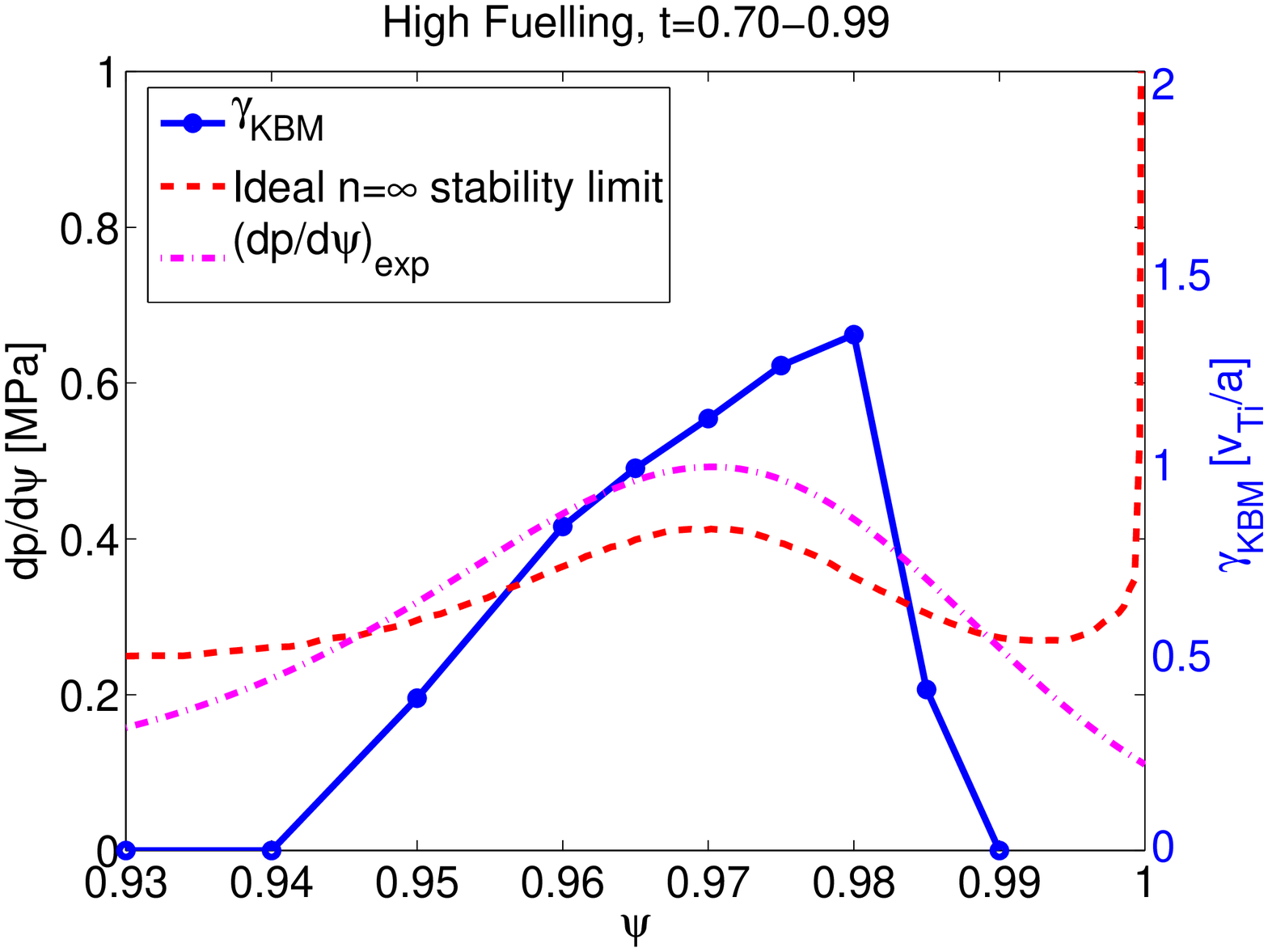,width=\linewidth}
\end{minipage}
\begin{minipage}{.5\linewidth}
\epsfig{file=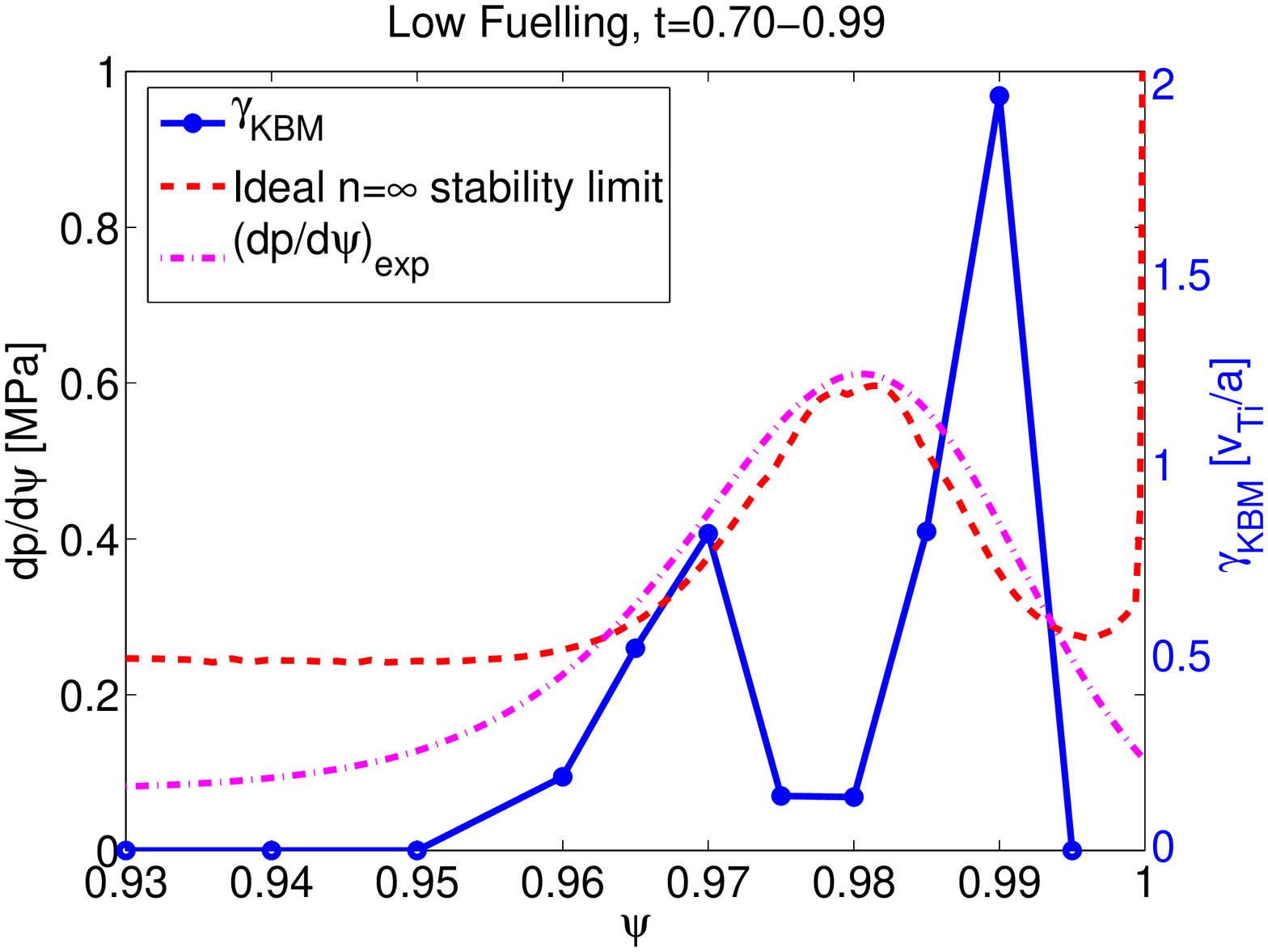,width=\linewidth}
\end{minipage}
\caption{The growth rate of the kinetic ballooning mode (solid blue
 line, y-axis on the right) at the end of the ELM cycle in the edge region
 of high (left) and low (right) fuelling case with no bootstrap current
 used in the equilibrium reconstruction. Also shown is the $n=\infty$
 ballooning stability limit (dashed red line, y-axis on the left) and the
 normalized pressure gradient (dash-dot magenta line, y-axis on the
 left).}\label{kbm_bs}
\end{figure}

\begin{figure}
\begin{minipage}{.5\linewidth}
\epsfig{file=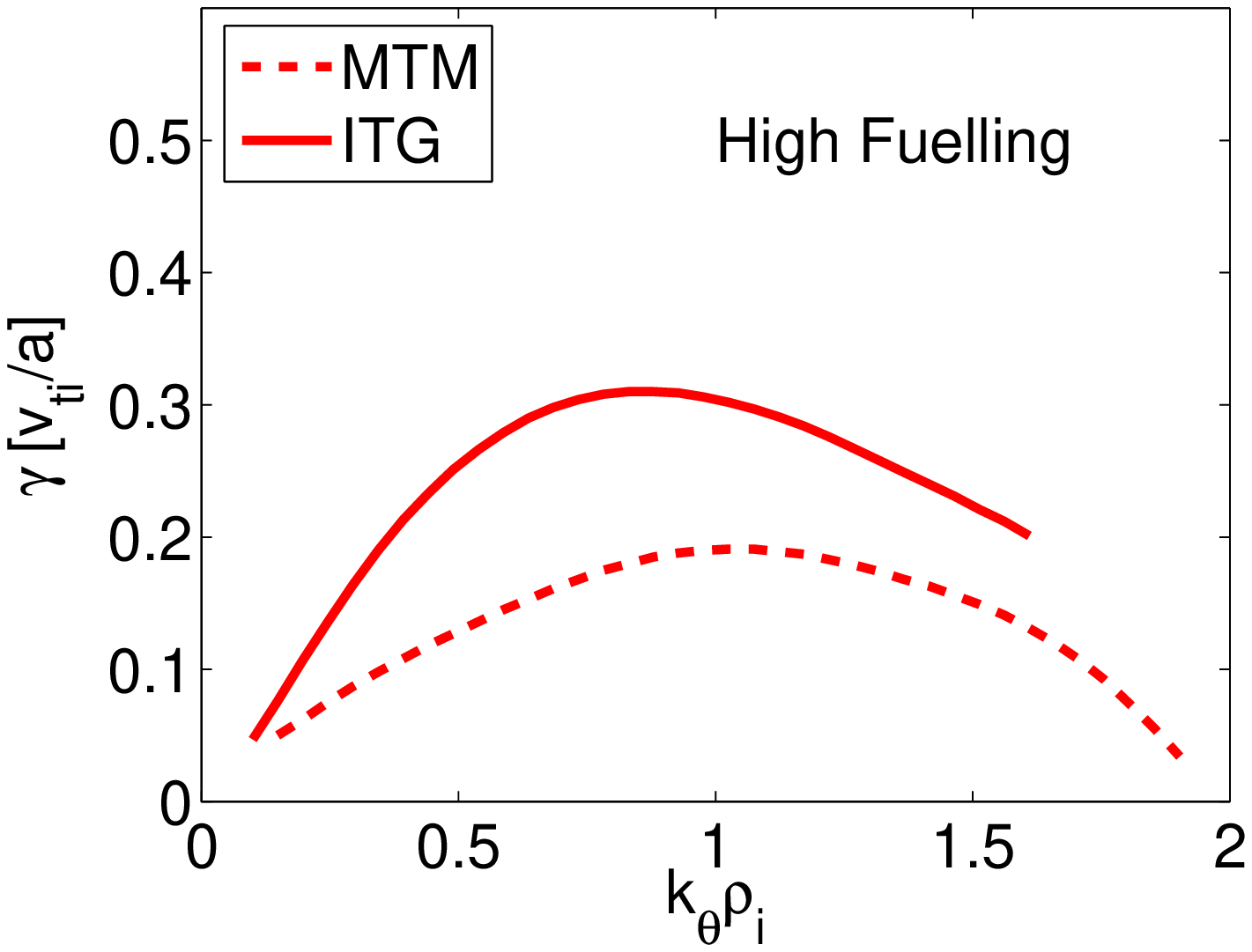,width=\linewidth}
\end{minipage}
\begin{minipage}{.5\linewidth}
\epsfig{file=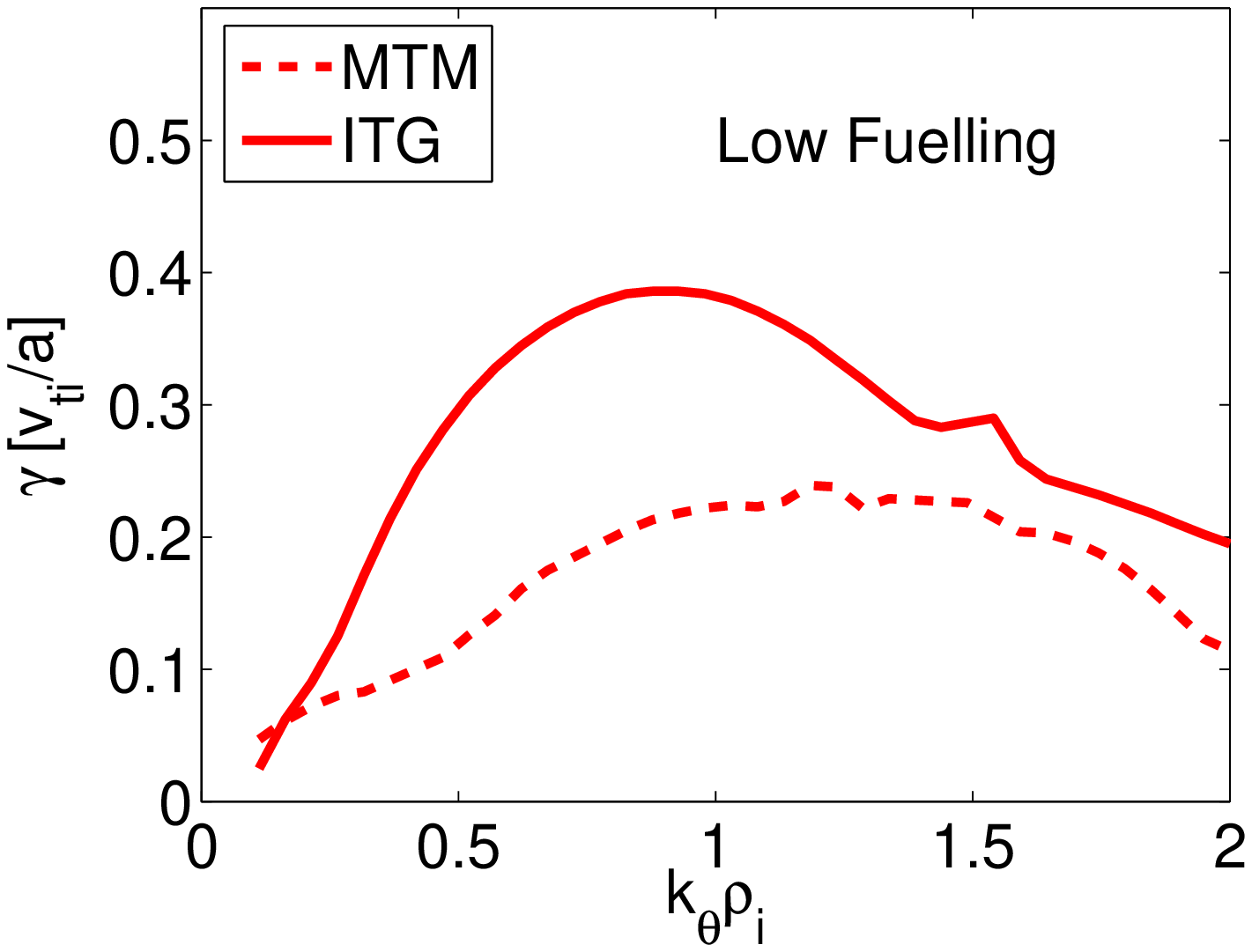,width=\linewidth}
\end{minipage}
\caption{The growth rate spectrum of MTM and ITG modes in high (left) and low (right) fuelling case at the
pedestal ``knee'' at the end of the ELM cycle. \added{ The ITG growth
rates are calculated without restrictions for growing modes, while the MTM
growth rates are obtained by suppressing all even parity modes.}}\label{fig:mutearitg}

\end{figure}
\begin{figure}
\begin{minipage}{.5\linewidth}
\epsfig{file=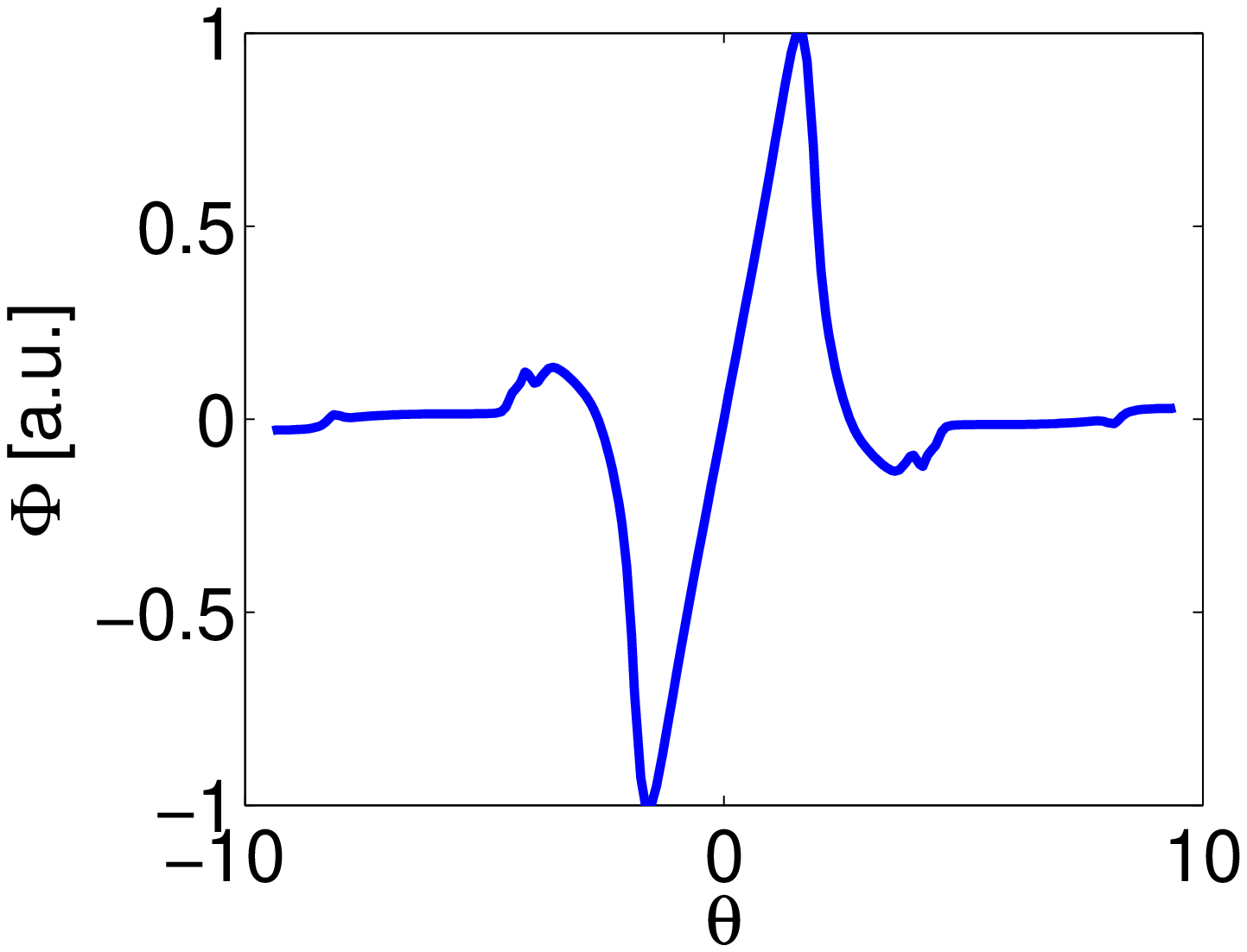,width=\linewidth}
\end{minipage}
\begin{minipage}{.5\linewidth}
\epsfig{file=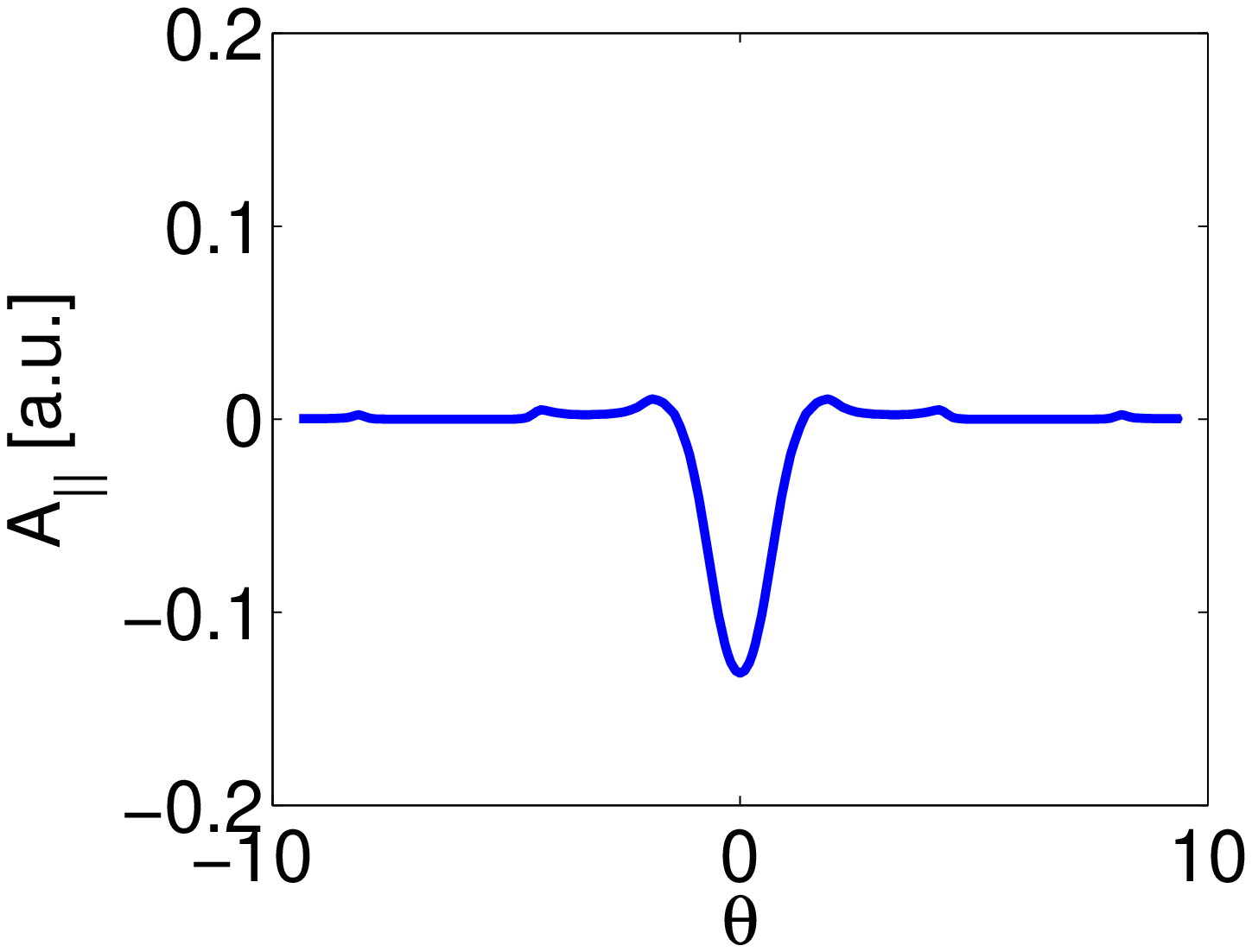,width=\linewidth}
\end{minipage}
\caption{The eigenfunctions of electrostatic potential $\phi$ and parallel
magnetic vector potential $A_{||}$ along the ballooning angle $\theta$ for
  the $k_y\rho_i=0.37$ micro-tearing mode at the JET pedestal
  top.}\label{fig:muteareigenfunc}
\end{figure}

\end{document}